\documentclass[pra,notitlepage,superscriptaddress,showpacs]{revtex4-1}
\usepackage[T1]{fontenc}
\usepackage{textcomp}
\usepackage{amsmath}
\usepackage{amssymb}
\usepackage{graphicx}
\usepackage{esint}

\makeatletter
\renewcommand{\textendash}{--}

\makeatother

\begin{document}

\title{Azimuthal modulation of electromagnetically induced transparency
using structured light}

\author{Hamid Reza Hamedi}
\email{hamid.hamedi@tfai.vu.lt}

\affiliation{Institute of Theoretical Physics and Astronomy, Vilnius University,
Saul\.etekio 3, Vilnius LT-10222, Lithuania}

\author{Viaceslav Kudria\v{s}ov}
\email{viaceslav.kudriasov@ff.vu.lt }

\affiliation{Institute of Theoretical Physics and Astronomy, Vilnius University,
Saul\.etekio 3, Vilnius LT-10222, Lithuania}

\author{Julius Ruseckas}
\email{julius.ruseckas@tfai.vu.lt}

\affiliation{Institute of Theoretical Physics and Astronomy, Vilnius University,
Saul\.etekio 3, Vilnius LT-10222, Lithuania}

\author{Gediminas Juzeli\=unas}
\email{gediminas.juzeliunas@tfai.vu.lt}

\affiliation{Institute of Theoretical Physics and Astronomy, Vilnius University,
Saul\.etekio 3, Vilnius LT-10222, Lithuania}
\begin{abstract}
Recently a scheme has been proposed for detection of the structured
light by measuring the transmission of a vortex beam through a cloud
of cold rubidium atoms with energy levels of the $\Lambda$-type configuration
{[}N. Radwell et al., Phys. Rev. Lett. 114, 123603 (2015){]}. This
enables observation of regions of spatially dependent electromagnetically
induced transparency (EIT). Here we suggest another scenario for detection
of the structured light by measuring the absorption profile of a weak
nonvortex probe beam in a highly resonant five-level combined tripod
and $\Lambda$ (CTL) atom-light coupling setup. We demonstrate that
due to the closed-loop structure of CTL scheme, the absorption of
the probe beam depends on the azimuthal angle and orbital angular
momentum (OAM) of the control vortex beams. This feature is missing
in simple $\Lambda$ or tripod schemes, as there is no loop in such
atom-light couplings. One can identify different regions of spatially
structured transparency through measuring the absorption of probe
field under different configurations of structured control light. 
\end{abstract}

\maketitle

\section{Introduction}

There has been a substantial interest in coherent control of the optical
properties of the medium in the last decades \cite{2005-RMP-Fleischhauer,2005-JPB-Andre,Fleischhauer2000}.
One remarkable manifestation of a coherently driven medium is the
phenomenon of electromagnetically induced transparency (EIT) \cite{1997-PD-Harris}.
EIT is a quantum interference effect where the destructive interference
between probability amplitudes of two optical transitions leads to
the elimination of absorption and an associated steep variation of
the refractive index around the resonant frequency \cite{1990-PRL-Harris,1991-PRL-Boller,1997-PD-Harris}.
This happens as a result of the dramatic modification of the optical
response for a weak probe field when the medium is simultaneously
exposed to another, strong laser field. Such a regime of coherent
light-matter interaction has led to the discovery of a number of fascinating
phenomena: slow and ultraslow light \cite{1999-Nature-Hau,1999-PRL-Budker},
light storage and retrieval \cite{2001-PRL-Phillips,2001-Nature-Liu},
stationary light \cite{2003-Nature-Bajcsy} and giant optical nonlinearities
\cite{1996-OL-Schmidt,2006PackKerr}. It was also demonstrated that
EIT allows for the coherent manipulation of individual photons and
efficient conversion of their quantum states into long-lived atomic
coherences \cite{2005-Nature-Eisaman,2005-Nature-Chaneliere}. The
latter is a desirable feature for all-optical quantum information
processing and quantum memory applications \cite{2017-JO-Ma,2003-RMP-Lukin}.

In its simplest form EIT requires only three atomic levels and two
light beams in a suitable atomic medium with configuration geometry
of the $\Lambda$ type \cite{2005-JPB-Andre,2017-JO-Ma,Fleischhauer2005}.
In general, however, EIT is not restricted to this simple scheme,
but takes place also in more complex atomic configurations where multiple
energy states interact with multiple laser fields \cite{2001-OC-McGloin,2002PRA-Paspalakis,2002-JOB-Paspalakis,2003-JPB-McGloin,2004-PLA-Wang,2007-JPB-Bhattacharyya,2017-JPB-Hamedi,2010-NC-Unanyan,Fleischhauer2016,LeeNatcom2014}.
Such complex EIT systems possess a range of useful features, like
multiple transparency windows, support of slow light at different
frequencies, and the diversity in the optical response characteristics.
Typically, the introduction of additional driving fields or atomic
levels allows better flexibility and control over the EIT. For example,
by applying a microwave field to the lower levels of the $\Lambda$
scheme, enhancement or suppression of EIT can be observed \cite{2009-PRA-Li}.
On the other hand, by adding an extra optical field to the $\Lambda$
scheme the group velocity of a probe pulse can be effectively manipulated
\cite{2001-PRA-Agarwal}. Despite numerous studies on EIT, there is
still a considerable interest in developing more advanced atom-light
coupling schemes and investigation of the associated new properties.

An increased attention has turned recently to the usage of the special
optical beams in EIT, particularly optical vortices \cite{2015-OC-Akin}.
An optical vortex is a beam with nonzero orbital angular momentum
(OAM), meaning that its optical phase is a function of azimuthal coordinate
and the wavefront is helical \cite{1992-PRA-Allen,2011-AOP-Yao}.
Some important features of OAM in the EIT regime have been demonstrated
including resonance narrowing \cite{2013-OC-Chanu} or the robustness
of stored OAM state to decoherence \cite{2007-PRL-Pugatch}, suggesting
vortex usability in the EIT-related applications. Moreover, OAM serves
as an additional degree of freedom for a photon and hence represents
a system of a higher dimension for the high capacity information transmission.
Quite recently, the OAM-based long distance communication over multi-kilometer
ranges have been demonstrated \cite{2016-PNAS-Krenn}, and quantum
entanglement of photons carrying high order topological charges have
been reported \cite{2016-PNAS-Fickler}. Combining the increased information
capacity of the vortex beams with the EIT-based possibilities of coherent
photon manipulation and trapping, this concept becomes a promising
approach for the quantum information science \cite{2013-OL-Veissier}.

One may take advantage of such a combined OAM and EIT approach to
study the characteristics of OAM in the atomic ensembles \cite{2011-PRA-Ruseckas,2013-PRA-Ruseckas}.
To this end, a couple of recent studies have demonstrated how the
coherent properties of EIT make it possible to identify the OAM information
by producing spatially variable absorption patterns \cite{2012-EPL-Han,2015-PRL-Radwell}.
It was suggested to identify the OAM by converting its spatial phase
information to the corresponding intensity distribution using a microwave
field between the ground levels in the $\Lambda$ scheme \cite{2012-EPL-Han}.
In a remarkable work by Radwell \emph{et al}.\ \cite{2015-PRL-Radwell},
the formation of the spatially varying EIT was demonstrated using
light beam with both azimuthally varying polarization and a phase
structure. In this work the existence of a spatially varying dark
state was shown at specific angles with the symmetry determined by
the input light polarization. However, the proposal dealt with a simple
three-level atom-light coupling scheme of the $\Lambda$-type which
has obvious limitations. Moreover, this study concentrated on situation
where the incident probe beam carries the OAM.

Here we suggest another scenario for the detection of structured light
patterns and formation of spatially varying optical transparency based
on a more complex combined tripod and $\Lambda$ (CTL) atom-light
coupling scheme (see Fig.~\ref{fig:scheme}(a)). Note that the spatially
varying optical transparency reported in \cite{2015-PRL-Radwell}
is achieved by exposing a three-level $\Lambda$-type atomic scheme
to a single light beam with an azimuthally varying polarization and
phase structure. The left and right-handed circular polarization components
form the probe and control fields for the EIT transition via a Hanle
resonance \cite{1997-PRA-Renzoni}, resulting in an azimuthal variation
of the dynamics of the atomic ensemble. On the other hand, here we
consider a situation in which the incident probe beam does not have
an optical vortex, whereas the control fields carry optical vortices.
We demonstrate that due to the closed-loop structure of the CTL scheme,
the probe absorption depends on the azimuthal angle and the OAM of
the control vortex beams. This feature is missing in simple $\Lambda$
or tripod schemes as there is no loop in such atom-light couplings.
In that case, even if the control fields carry optical vortices $\propto\exp(il\Phi)$
($\Phi$ and $l$ being the azimuthal angle and OAM number, respectively),
the probe field does not feel the effect induced by vortex control
beams as the magnitude squared of control fields appears in expression
for the probe susceptibility \cite{2002PRA-Paspalakis,2002-JOB-Paspalakis}.
We show that a more complex atomic setup made by a combination of
two $\Lambda$ and tripod schemes can be utilized to overcome this
downside enabling to measure the regions of optical transparency.
Such a model may provide a promising approach to identify the OAM
of control fields by mapping the spatially dependent absorption profile
of the probe field.

\section{Model and formulation \label{sec:Model-and-formulation}}

\begin{figure}
\includegraphics[width=0.3\columnwidth]{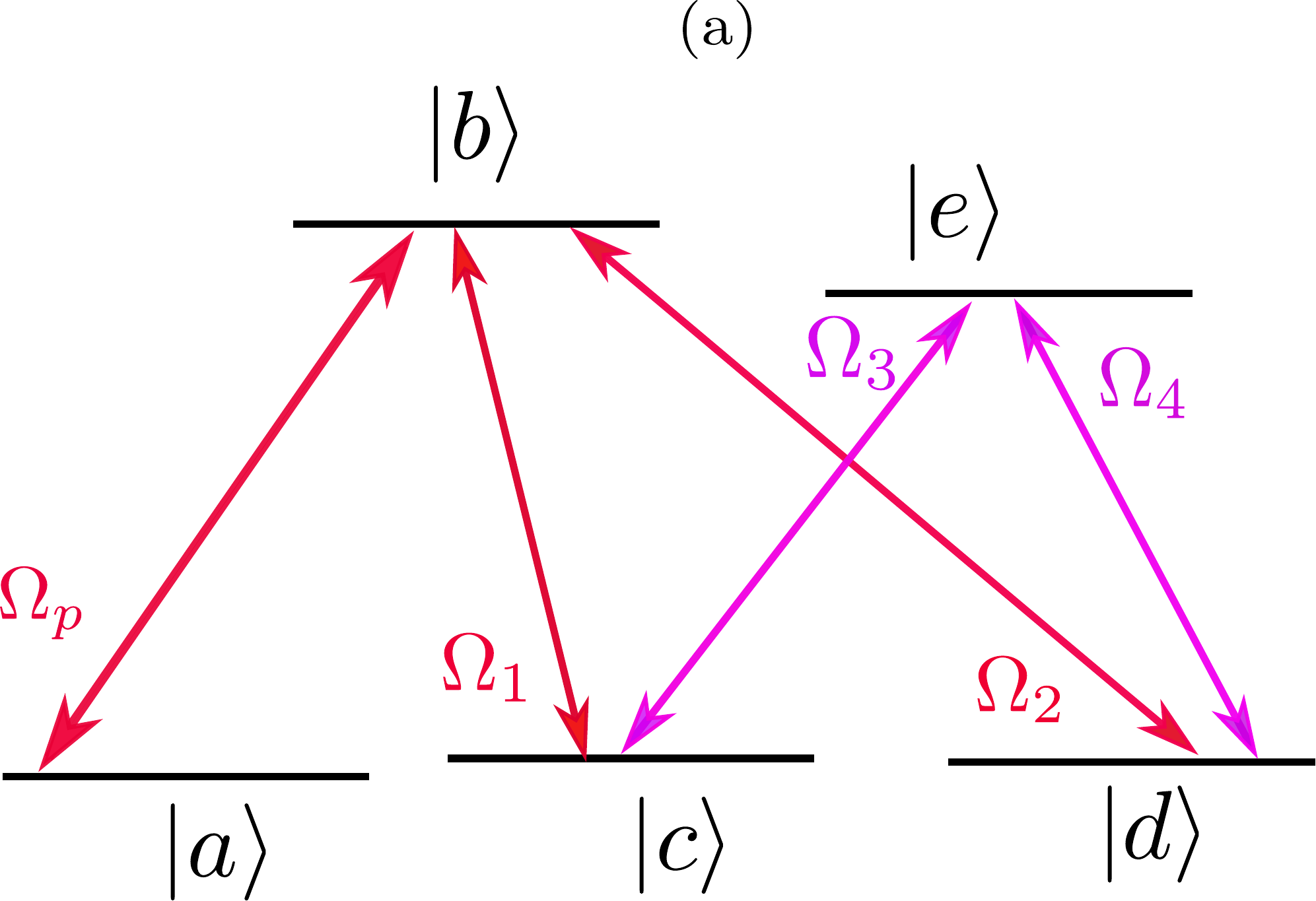} \includegraphics[width=0.3\columnwidth]{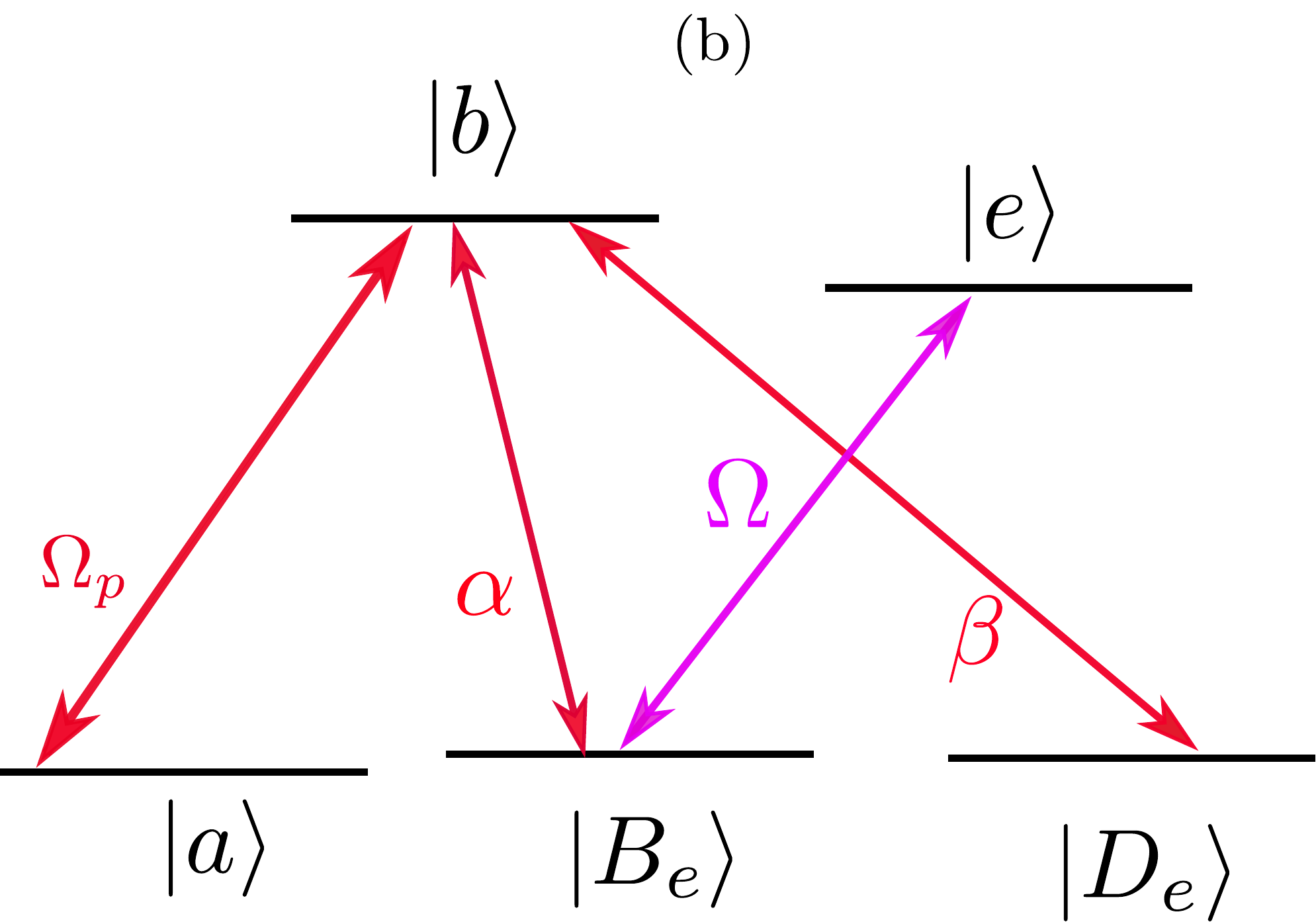}
\caption{Five-level combined tripod and $\Lambda$ atomic system (a). Five-level
combined tripod and $\Lambda$ atomic system in the transformed basis
for $\alpha,\beta\protect\neq0$ (b).}
\label{fig:scheme}
\end{figure}

We consider a five-level CTL atom-light coupling scheme shown in Fig.~\ref{fig:scheme}(a).
The scheme is made of a four-level tripod subsystem (consisting of
atomic levels $|a\rangle$, $|b\rangle$, $|c\rangle$, and $|d\rangle$)
as well as a three-level $\Lambda$ subsystem (including atomic levels
$|c\rangle$, $|e\rangle$, and $|d\rangle$) coherently coupled to
each other by a probe field and four control laser fields. The probe
and control fields are assumed to co-propagate along the same direction.
Four control laser fields represented by Rabi frequencies $\Omega_{1},$
$\Omega_{2}$, $\Omega_{3}$ and $\Omega_{4}$ couple two excited
states $|b\rangle$ and $|e\rangle$ via two different pathways $|b\rangle\stackrel{\Omega_{1}^{*}}{\rightarrow}|c\rangle\stackrel{\Omega_{3}}{\rightarrow}|e\rangle$
and $|b\rangle\stackrel{\Omega_{2}^{*}}{\rightarrow}|d\rangle\stackrel{\Omega_{4}}{\rightarrow}|e\rangle$
making a four-level closed-loop coherent coupling scheme described
by the Hamiltonian ($\hbar=1$)

\begin{equation}
H_{\mathrm{4Levels}}=-\Omega_{1}^{*}|c\rangle\langle b|-\Omega_{2}^{*}|d\rangle\langle b|-\Omega_{3}^{*}|c\rangle\langle e|-\Omega_{4}^{*}|d\rangle\langle e|+\mathrm{H.c.}\,.\label{eq:1}
\end{equation}
A weak probe field described by a Rabi frequency $\Omega_{p}$ couples
then the closed-loop subsystem to a ground level $|a\rangle$ via
the atomic transition $|a\rangle\longleftrightarrow|b\rangle$. The
destructive interference between different transition pathways induced
by the control and probe beams can make the medium transparent for
the resonant probe beam in a narrow frequency range due to the EIT
\cite{2017-JPB-Hamedi}. The total Hamiltonian of the system involving
all five atomic levels of the CTL level scheme is described by 
\begin{equation}
H_{\mathrm{5Levels}}=-\left(\Omega_{p}^{*}|a\rangle\langle b|+\Omega_{p}|b\rangle\langle a|\right)+H_{\mathrm{4Levels}}\,.\label{eq:2}
\end{equation}

In a new basis, the Hamiltonian for the atomic four-level subsystem
(\ref{eq:1}) can be expressed as \cite{2017-JPB-Hamedi}

\begin{equation}
H_{\mathrm{4Levels}}=-\beta|D_{e}\rangle\langle b|-\alpha|B_{e}\rangle\langle b|-\Omega|B_{e}\rangle\langle e|+\mathrm{H.c.},\label{eq:3}
\end{equation}
where
\begin{align}
|D_{e}\rangle & =\frac{1}{\Omega}\left(\Omega_{4}|c\rangle-\Omega_{3}|d\rangle\right),\label{eq:D_e}\\
|B_{e}\rangle & =\frac{1}{\Omega}\left(\Omega_{3}^{*}|c\rangle+\Omega_{4}^{*}|d\rangle\right),\label{eq:B_e}
\end{align}
are the internal dark and bright states for the $\Lambda$-scheme
made of the two ground states states $|c\rangle$ and $|d\rangle$,
as well as an excited states $|e\rangle$. In writing Eq.~(\ref{eq:3}),
we define
\begin{align}
\beta & =\frac{1}{\Omega}(\Omega_{1}^{*}\Omega_{4}^{*}-\Omega_{2}^{*}\Omega_{3}^{*}),\label{eq:beta}\\
\alpha & =\frac{1}{\Omega}(\Omega_{1}^{*}\Omega_{3}+\Omega_{2}^{*}\Omega_{4}),\label{eq:alpha}
\end{align}
and the total Rabi frequency 
\begin{equation}
\Omega=\sqrt{|\Omega_{3}|^{2}+|\Omega_{4}|^{2}}\,.\label{eq:Omega}
\end{equation}
By changing the coefficients $\beta$ and $\alpha$ one arrives at
three different situations: i) both $\alpha$ and $\beta$ are nonzero;
ii) $\beta$ is zero; iii) $\alpha$ is zero \cite{2017-JPB-Hamedi}.
When both $\alpha$ and $\beta$ are nonzero ($\alpha,\beta\neq0$),
there exists a superposition state 

\begin{equation}
|D\rangle=\beta|a\rangle-\Omega_{p}|D_{e}\rangle,\label{eq:Dark}
\end{equation}
which has no contribution from both bare excited states $|b\rangle$
and $|e\rangle$. Assuming that the probe field is much weaker than
the control fields, it follows from Eq.~(\ref{eq:Dark}) that the
superpositon state $|D\rangle$ is approximately equal to $\beta|a\rangle$,
indicating that the ground state $|a\rangle$ becomes a dark state
and not interact with the light forming a transparency window around
the zero probe detuning $\Delta_{p}$.

\section{Vortex dependent probe absorption \label{sec:vort-dep-susc}}

In the CTL scheme, the dynamics of the probe field and the atomic
coherences can be described by the optical Bloch equations. The reduced
optical Bloch equations in the new basis are 

\begin{align}
\dot{\rho}_{ba} & =-(\Gamma_{b}/2-i\Delta_{p})\rho_{ba}+i\alpha\rho_{B_{e}a}+i\beta\rho_{D_{e}a}+i\Omega_{p},\label{eq:4}\\
\dot{\rho}_{B_{e}a} & =i\Delta_{p}\rho_{B_{e}a}+i\alpha^{*}\rho_{ba}+i\Omega^{*}\rho_{ea},\label{eq:5}\\
\dot{\rho}_{D_{e}a} & =i\Delta_{p}\rho_{D_{e}a}+i\beta^{*}\rho_{ba},\label{eq:6}\\
\dot{\rho}_{ea} & =-(\Gamma_{e}/2-i\Delta_{p})\rho_{ea}+i\Omega\rho_{B_{e}a},\label{eq:7}
\end{align}
where $\rho_{ba}$ is the optical coherence corresponding to the probe
transition of $|a\rangle\rightarrow|b\rangle$, while $\rho_{B_{e}a}$,
$\rho_{D_{e}a}$ or $\rho_{ea}$ are the ground-state coherence between
$|a\rangle$ and $|B_{e}\rangle$, $|D_{e}\rangle$ or $|e\rangle$.
Note that we have assumed that the probe field is much weaker than
the control ones. In that case most atomic population is in the ground
state, and one can treat the probe field as a perturbation. All fast-oscillating
exponential factors associated with central frequencies and wave vectors
have been eliminated from the equations, and only the slowly-varying
amplitudes are retained. Here we have defined the probe detuning as
$\Delta_{p}=\omega_{p}-\omega_{ba},$ where $\omega_{p}$ is a central
frequency of the probe field. The control fields are assumed to be
on resonance. Two excited states $|b\rangle$ and $|e\rangle$ decay
with rates $\Gamma_{b}$ and $\Gamma_{e}$, respectively.

As it is known, the imaginary part of $\rho_{ba}$ corresponds to
the probe absorption. From the reduced OB equations (\ref{eq:4})\textendash (\ref{eq:7})
it is easy to obtain the steady-state solution to the density matrix
element $\rho_{ba}$

\begin{equation}
\rho_{ba}=\Omega_{p}\frac{\Delta_{p}\left(-|\Omega|^{2}+i\Delta_{p}\left(\Gamma_{e}/2-i\Delta_{p}\right)\right)}{i\Delta_{p}\left(\Gamma_{e}/2-i\Delta_{p}\right)\zeta+i|\Omega|^{2}\Delta_{p}\left(\Gamma_{b}/2-i\Delta_{p}\right)+\left(\Gamma_{b}/2-i\Delta_{p}\right)\Delta_{p}^{2}\left(\Gamma_{e}/2-i\Delta_{p}\right)-|\Omega|^{2}|\beta|^{2}},\label{eq:linear}
\end{equation}
where $\zeta=|\alpha|^{2}+|\beta|^{2}=|\Omega_{1}|^{2}+|\Omega_{2}|^{2}$. 

Having obtained the coherence term $\rho_{ba}$, the analytical solution
for the probe absorption $\mathrm{Im}(\rho_{ba})$ can be easily obtained,
after some straightforward algebra

\begin{equation}
\mathrm{Im}(\rho_{ba})=\Omega_{p}\frac{|\Omega|^{2}B\Delta_{p}+\Delta_{p}^{2}A\Gamma_{e}/2-\Delta_{p}^{3}B}{A^{2}+B^{2}},\label{eq:absr}
\end{equation}
with 

\begin{equation}
A=-\Delta_{p}^{4}+\Delta_{p}^{2}(\zeta+|\Omega|^{2}+\Gamma_{e}\Gamma_{b}/4)-|\Omega|^{2}|\beta|^{2},\label{eq:A}
\end{equation}
and 

\begin{equation}
B=-\Delta_{p}^{3}(\Gamma_{e}+\Gamma_{b})/2+\Delta_{p}\left(|\Omega|^{2}\Gamma_{b}/2+\zeta\Gamma_{e}/2\right).\label{eq:coeffBB}
\end{equation}
One can see that the probe absorption for the CTL system has now a
term 

\begin{equation}
|\Omega|^{2}|\beta|^{2}=|\Omega_{1}|^{2}|\Omega_{4}|^{2}+|\Omega_{2}|^{2}|\Omega_{3}|^{2}-Q,\label{eq:beta^2}
\end{equation}
entering to the parameter $A$ featured in Eq.~(\ref{eq:A}), where

\begin{equation}
Q=\Omega_{1}^{*}\Omega_{2}\Omega_{3}\Omega_{4}^{*}+\Omega_{1}\Omega_{2}^{*}\Omega_{3}^{*}\Omega_{4},\label{eq:QIT}
\end{equation}
represents the quantum interference between the four control fields.

This indicates that if a single control field or a combination of
several control fields carry an optical vortex ($\sim\exp(il\Phi)$)
in the CTL scheme, the linear absorption of the probe field given
by Eq~(\ref{eq:absr}) depends on the OAM of the control field(s)
through the parameter $Q$. Note that this does not mean that the
probe beam can acquire the OAM of control field(s) when propagating
in such a medium. The incoming probe beam does not have any vortex
initially at the beginning of the atomic medium. However, due to the
closed-loop structure of the atomic system, the probe beam subsequently
develops some OAM features by propagating in such a highly resonant
medium, making the transverse profile of the probe beam spatially
varying. This does not necessary mean that one arrives at a pure vortex
beam. Although the probe field acquires some non-zero OAM components
along with the zero OAM component, the intensity of the probe beam
does not go to zero at some transverse point. 

The origin of such sensitivity of the probe field to the OAM of the
control fields originates from the fact that the CTL system is phase-dependent
\cite{2017-JPB-Hamedi}. Such a phase sensitivity is contained in
the parameter $Q$ featured in Eq.~(\ref{eq:QIT}). Therefore, the
linear susceptibility and hence the probe absorption $\mathrm{Im}(\rho_{ba})$
will depend on the azimuthal angle $\Phi$ of the control fields carrying
the OAM. This can be exploited to measure the regions of spatially
varying transparency through measuring the linear absorption of probe
field. In addition, it may provide a promising approach to identify
the winding number of control fields by mapping the spatially dependent
absorption profile of the probe field.

\section{Azimuthal modulation of absorption profile \label{sec:azim-mod-abso}}

In last Section we demonstrated that the application of structured
lights make the probe absorption profile given by Eq.~(\ref{eq:absr})
spatially dependent due to the quantum interference term $Q$. This
indicates that by measuring the absorption profile of probe field
we can determine the regions of optical transparency or absorption.
Apparently, different cases of interaction of the atom with the vortex
control beams result in different absorption profiles so that different
patterns can be achieved for regions of spatially structured transparency.
In what follows we investigate the spatially dependent EIT for the
CTL system near the resonance (i.~e., $\Delta_{p}\ll\Gamma$) by
considering different cases of structured control light. We assume
$\Gamma_{e}=\Gamma_{b}=\Gamma$ and all parameters are scaled with
$\Gamma$ which should be of the order of MHz, like, for example,
in cesium (Cs) atoms. We will consider the following configurations
with increasing number of vortex beams: a) only the control field
$\Omega_{2}$ carries an optical vortex, the other control fields
$\Omega_{1},\Omega_{3}$ and $\Omega_{4}$ have no vortices; b) two
control fields $\Omega_{2}$ and $\Omega_{3}$ are vortex beams; c)
the control fields $\Omega_{3}$ and $\Omega_{4}$ carry optical vortices;
d) two control fields $\Omega_{2}$ and $\Omega_{4}$ are vortex beams;
e) three control fields $\Omega_{2}$, $\Omega_{3}$ and $\Omega_{4}$
are vortex beams; f) all four control fields carry optical vortices.
The amplitude of a vortex beam $|\Omega_{j}|$ is

\begin{equation}
|\Omega_{j}|=\epsilon_{j}\left(\frac{r}{w}\right)^{|l_{j}|}\exp\left(-\frac{r^{2}}{w^{2}}\right)\,,\label{eq:amplitude}
\end{equation}
where $l_{j}$ is an integer representing the vorticity, $r$ corresponds
to the distance from the vortex core (cylindrical radius) and $w$
stands for the beam waist parameter.

\begin{figure}
\includegraphics[width=0.3\columnwidth]{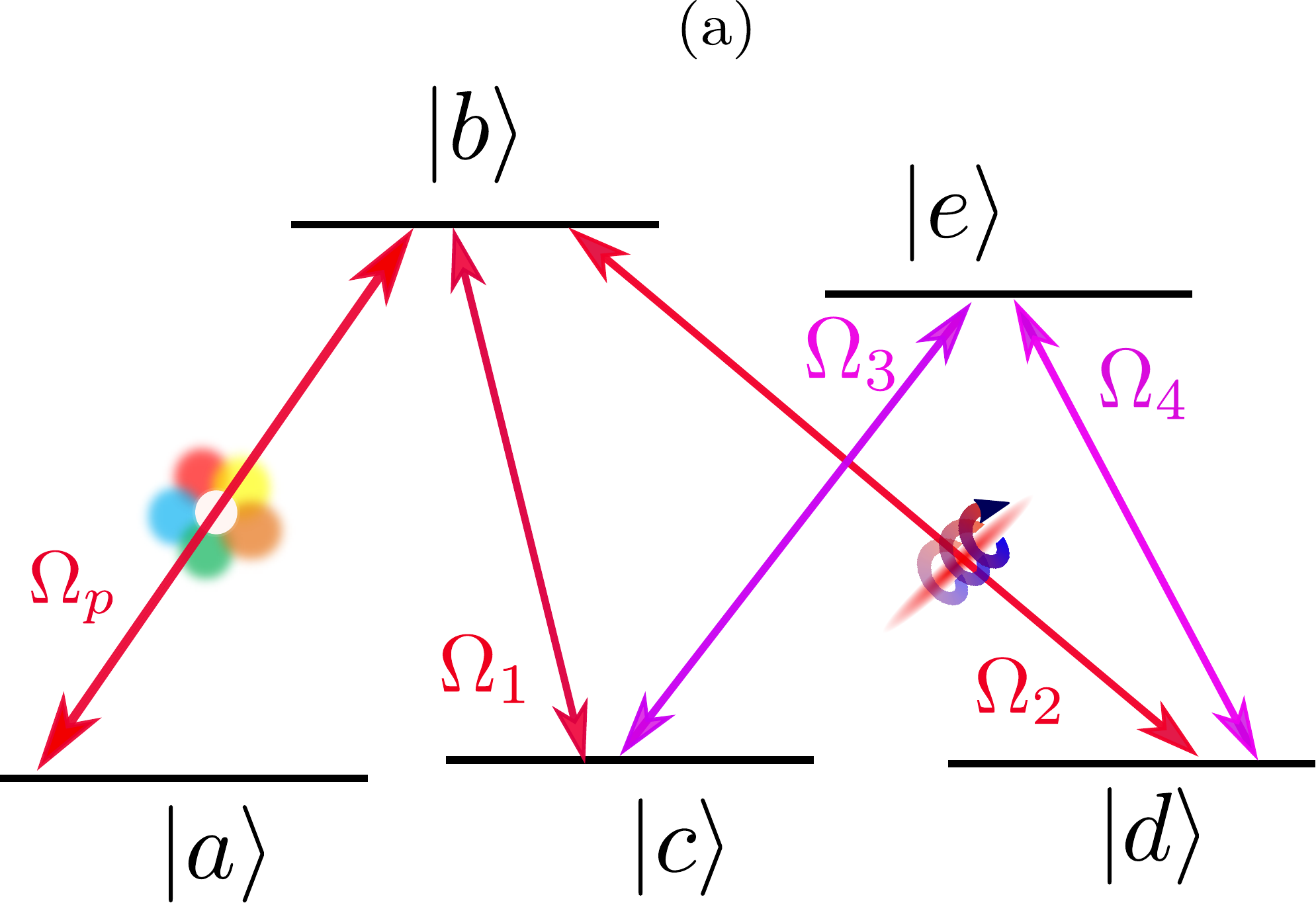} \includegraphics[width=0.3\columnwidth]{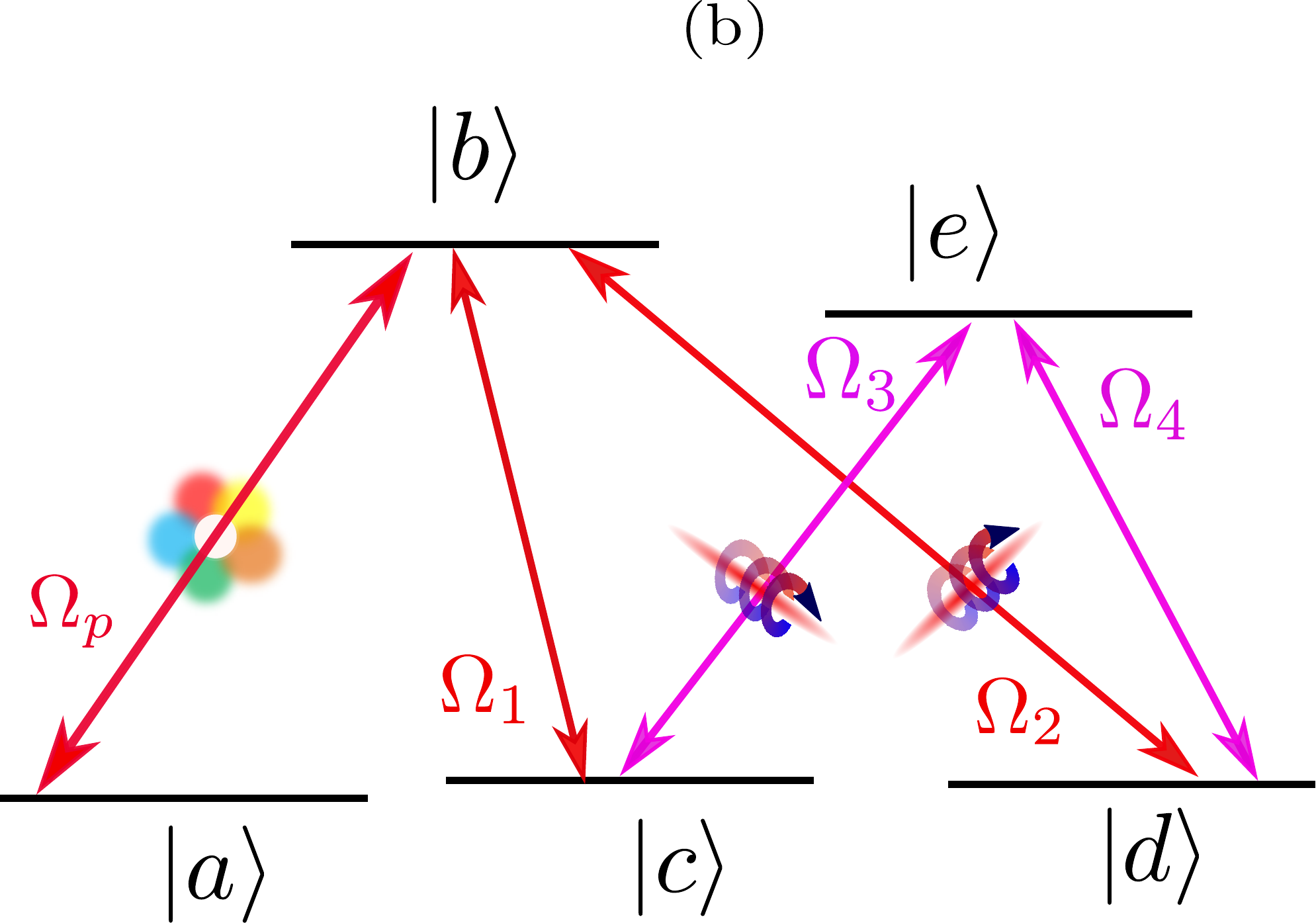}
\includegraphics[width=0.3\columnwidth]{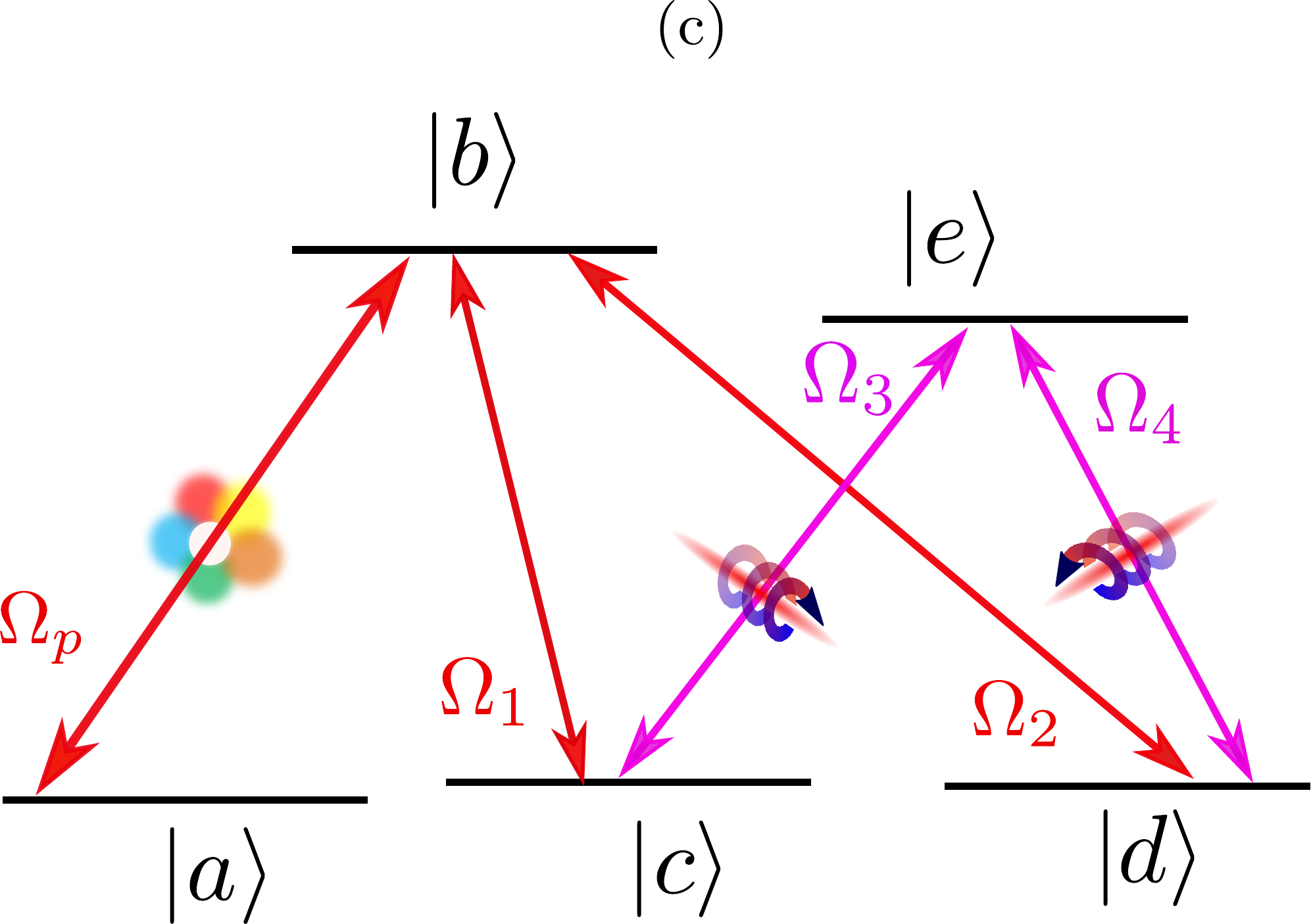} \includegraphics[width=0.3\columnwidth]{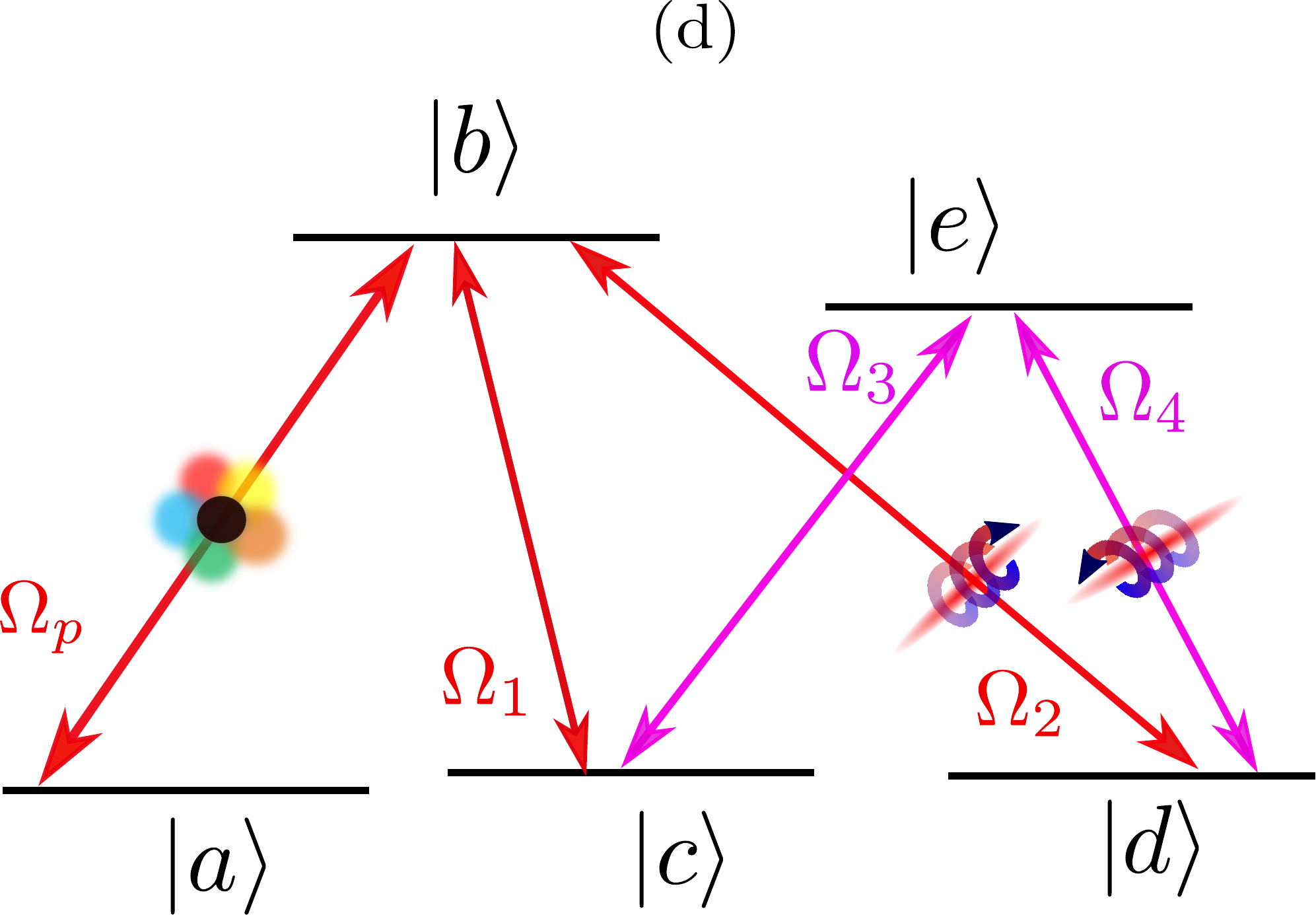}
\includegraphics[width=0.3\columnwidth]{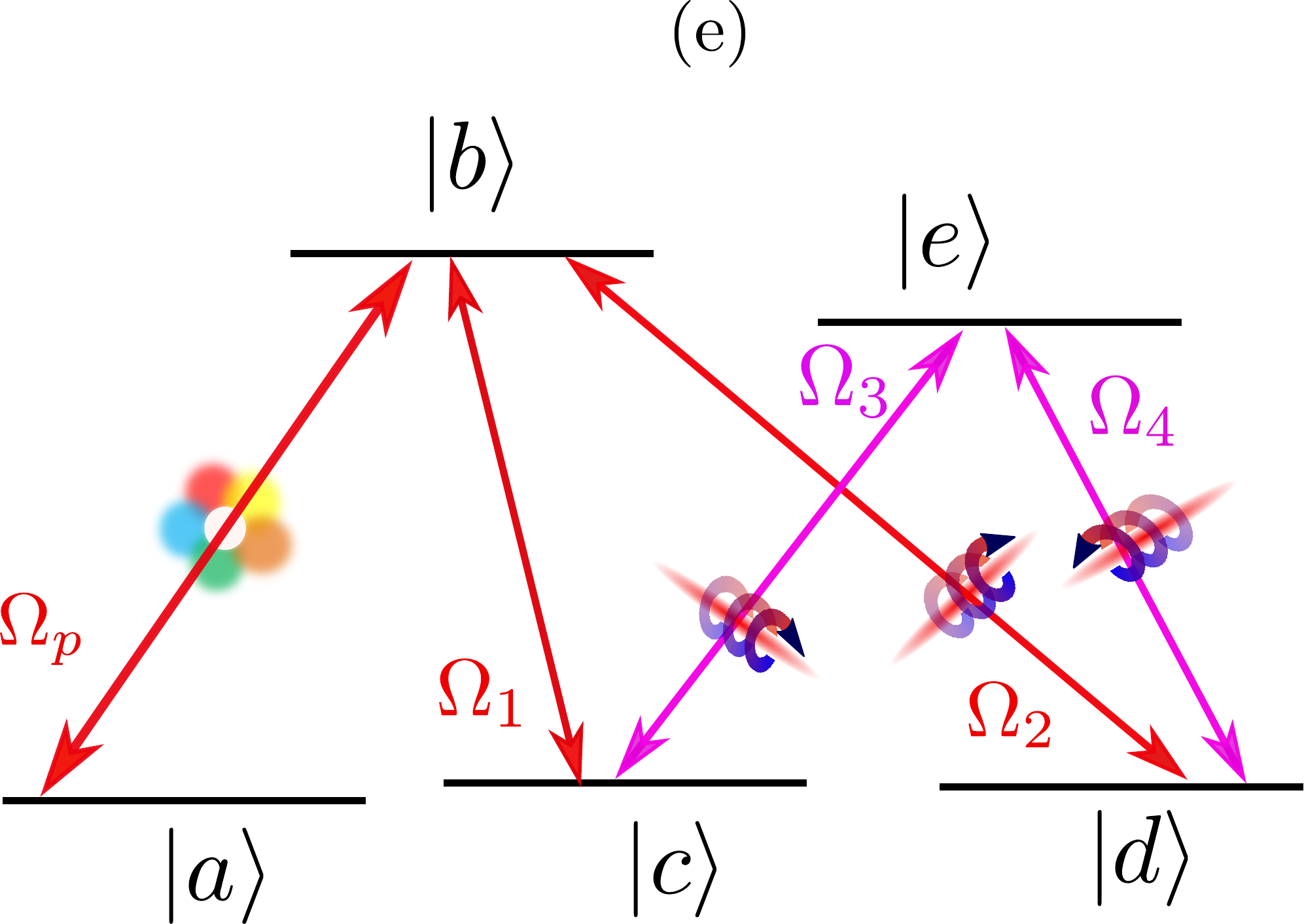} \includegraphics[width=0.3\columnwidth]{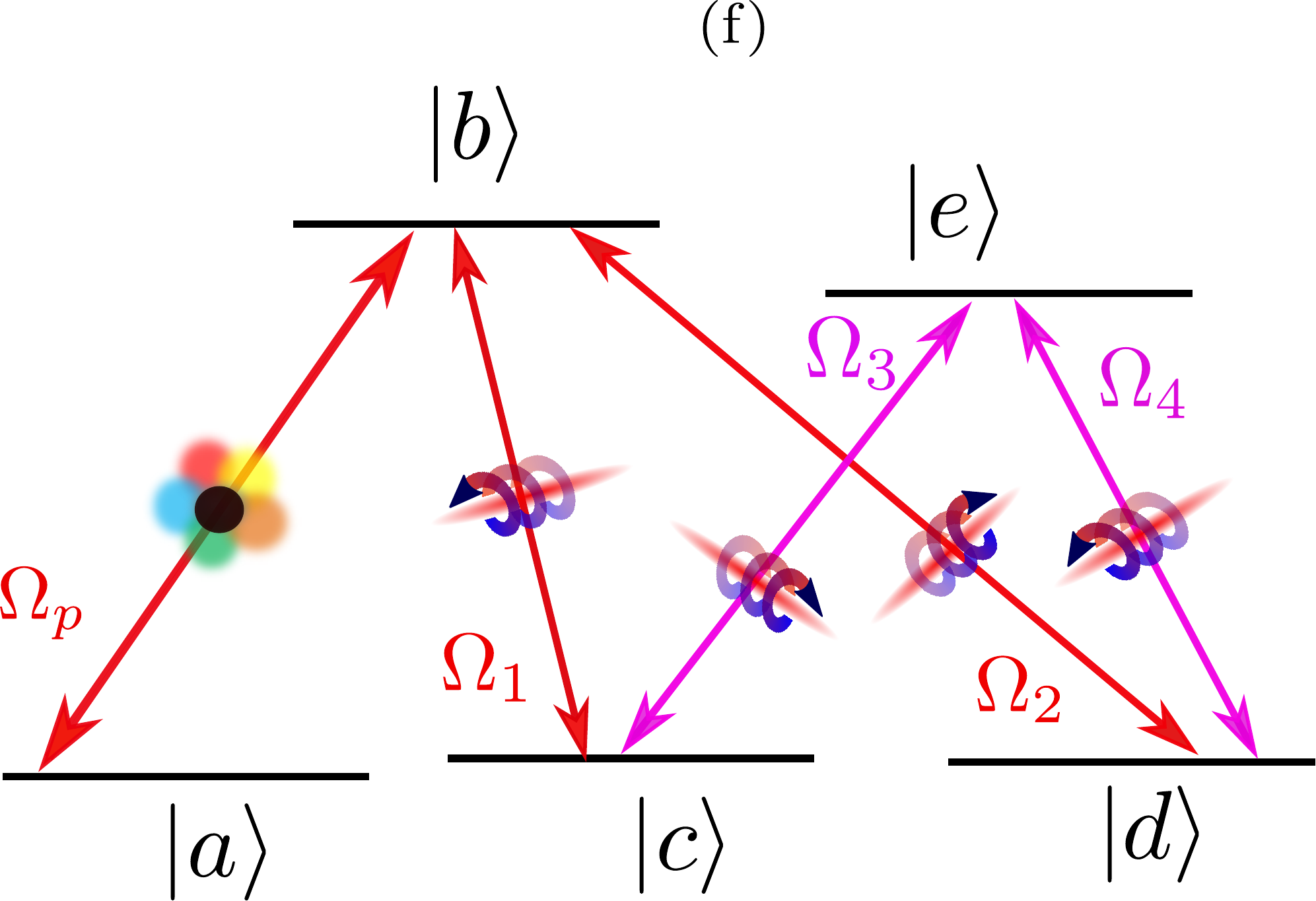} 

\caption{Different coupling scenarios of CTL atom with structured light, resulting
in different regions of spatially structured transparency. }
\label{fig:schemes}
\end{figure}

\subsubsection{Situation (a): single vortex beam}

At first let us consider a situation where the control field $\Omega_{2}$
carries an optical vortex $\Omega_{2}\backsim\exp(il_{2}\Phi)$, yet
other control fields $\Omega_{1},\Omega_{3}$ and $\Omega_{4}$ have
no vortices (Fig.~\ref{fig:schemes}(a)). Thus the Rabi-frequency
of vortex beam $\Omega_{2}$ is

\begin{equation}
\Omega_{2}=|\Omega_{2}|\exp(il_{2}\Phi),\label{eq:vortexBeam(a)2}
\end{equation}
where $\Phi$ is the azimuthal angle, $l_{2}$ is an integer representing
the vorticity of control field, while $|\Omega_{2}|$ is defined by
Eq.~\ref{eq:amplitude} with $j=2$. The other nonvortex control
field Rabi-frequencies are defined by

\begin{align}
\Omega_{1} & =|\Omega_{1}|\,,\label{eq:OtherBeams(a)1}\\
\Omega_{3} & =|\Omega_{3}|\,,\label{eq:OtherBeams(a)3}\\
\Omega_{4} & =|\Omega_{4}|\,.\label{eq:OtherBeams(a)4}
\end{align}
Under this situation, the quantum interference term defined by Eq.~(\ref{eq:QIT})
takes the form

\begin{equation}
Q=2|\Omega_{1}||\Omega_{2}||\Omega_{3}||\Omega_{4}|\cos(l_{2}\Phi).\label{eq:QI1}
\end{equation}

\begin{figure}
\includegraphics[width=0.3\columnwidth]{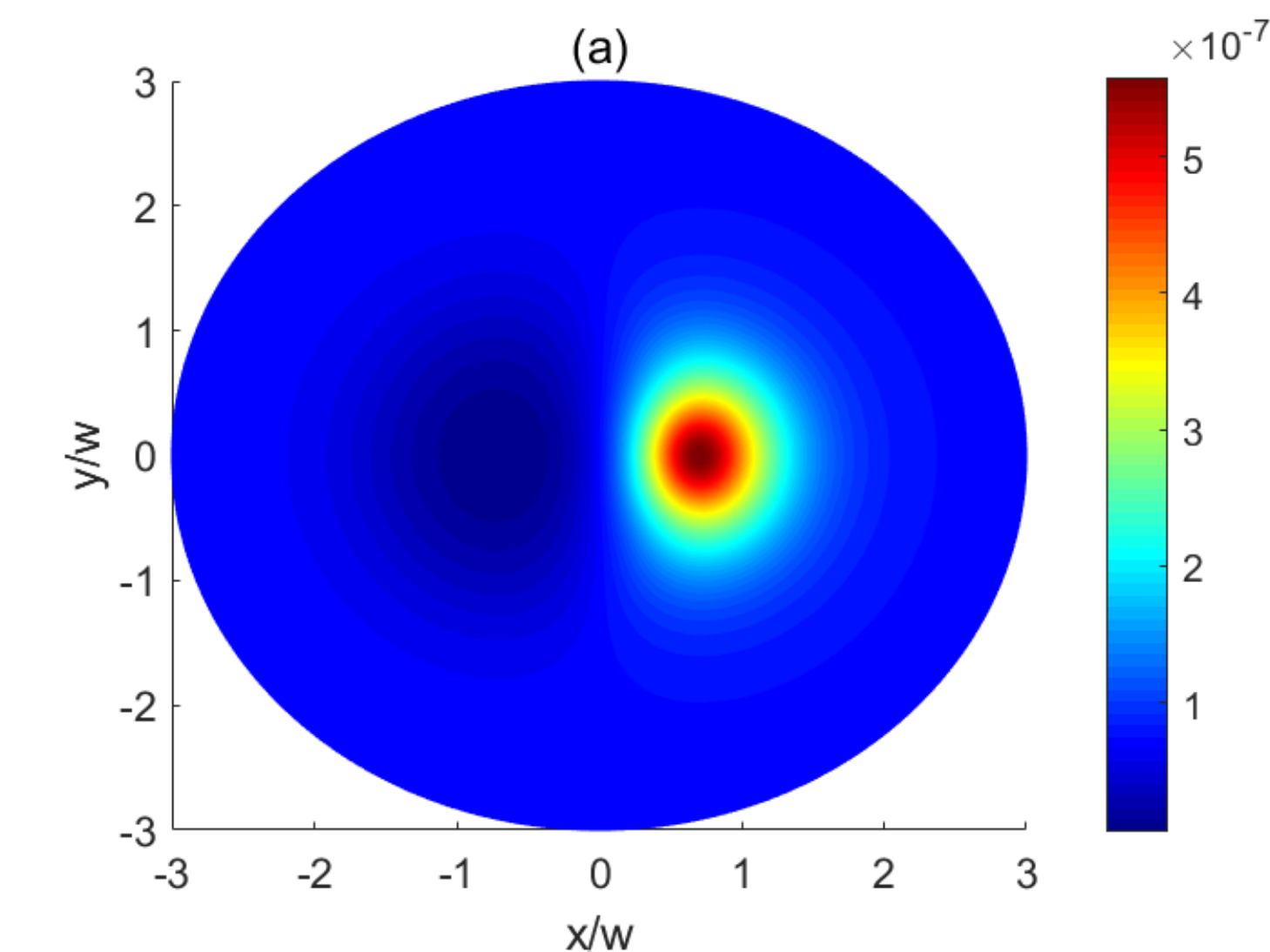} \includegraphics[width=0.3\columnwidth]{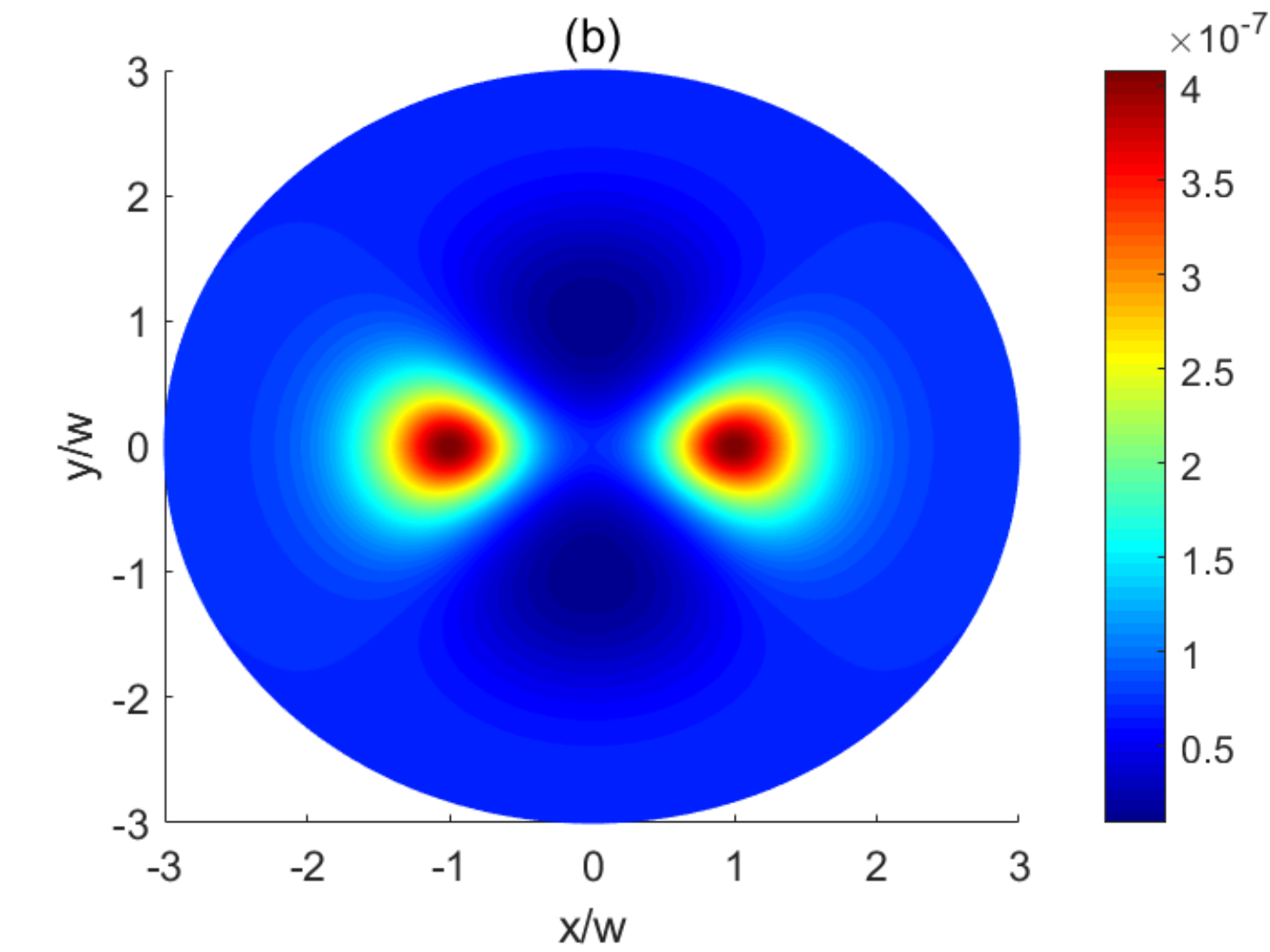}
\includegraphics[width=0.3\columnwidth]{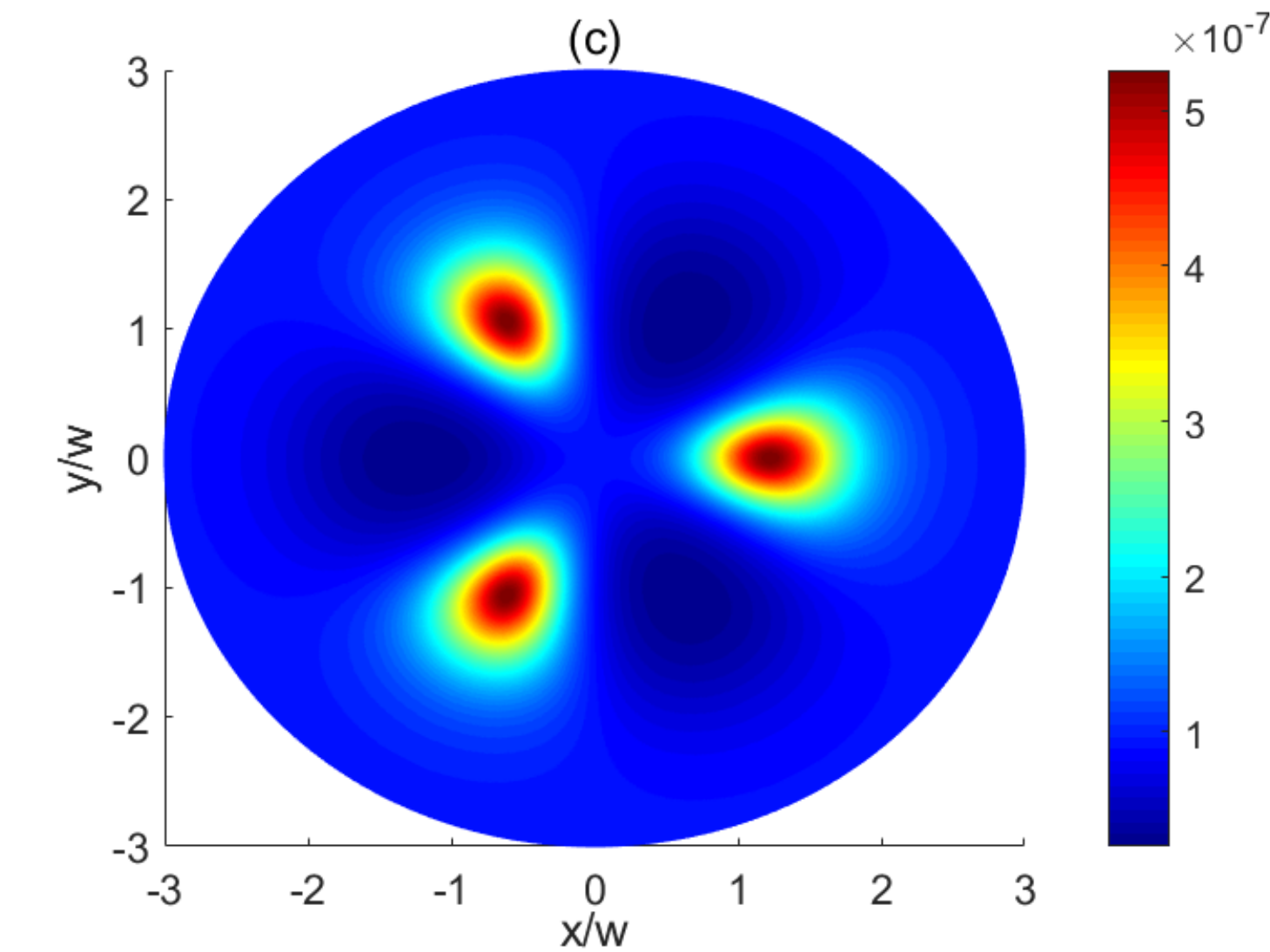}

\includegraphics[width=0.3\columnwidth]{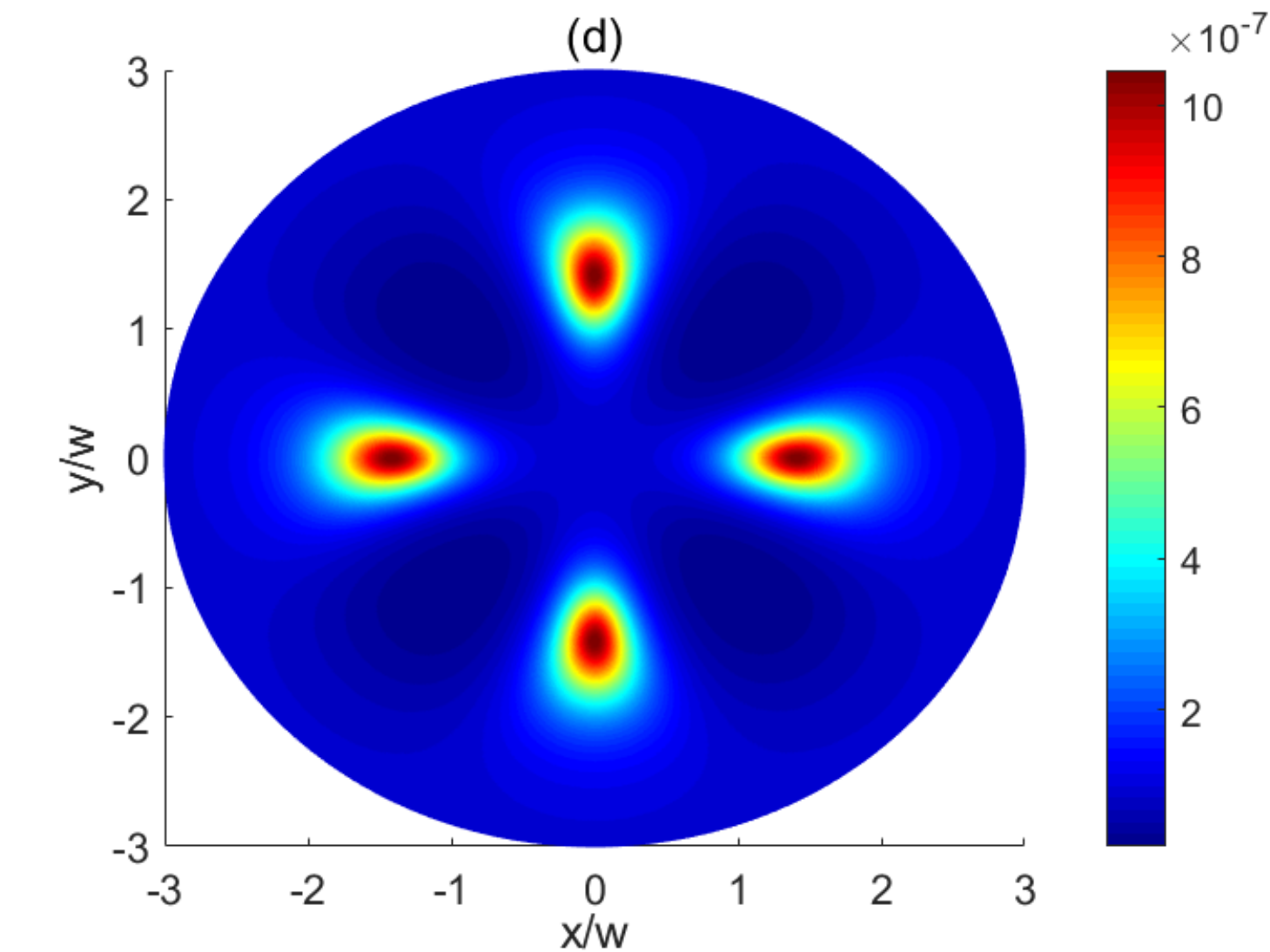} \includegraphics[width=0.3\columnwidth]{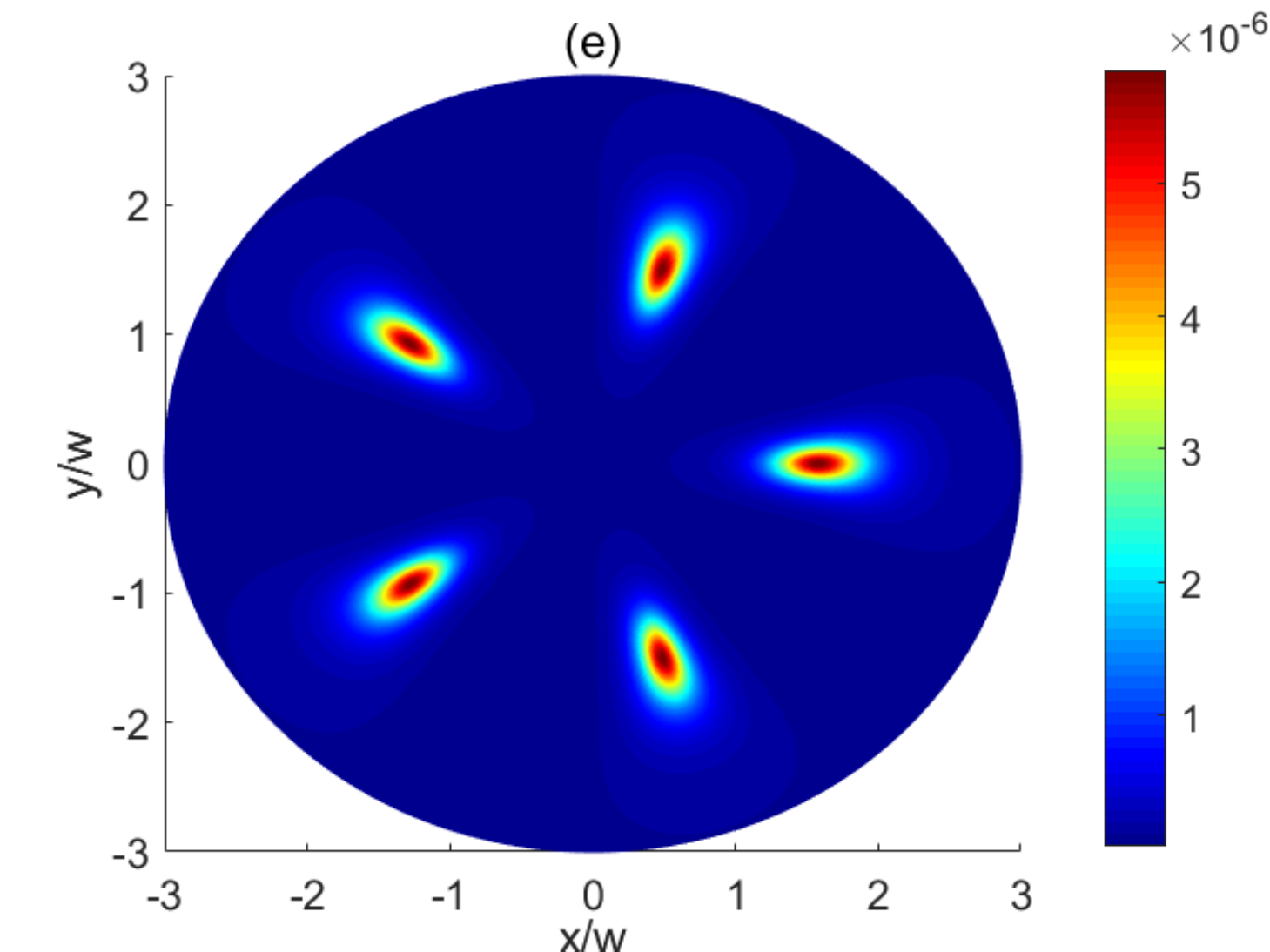}
\caption{Spatially dependent absorption profile of the probe beam in arbitrary
units when only the control field $\Omega_{2}$ has an optical vortex.
The vorticities are $l_{2}=1$ (a), $l_{2}=2$ (b), $l_{2}=3$ (c),
$l_{2}=4$ (d) and $l_{2}=5$ (e). Other parameters are $|\Omega_{1}|=0.6\Gamma,$
$\epsilon_{2}=0.7\Gamma,$ $|\Omega_{3}|=0.3\Gamma,$ $|\Omega_{4}|=0.5\Gamma,$
$\Gamma_{e}=\Gamma_{b}=\Gamma$ and $\Delta_{p}=0.001\Gamma$. }
\label{fig:fig3}
\end{figure}

Figure~\ref{fig:fig3} illustrates the absorption profile $\mathrm{Im}(\rho_{ba})$
based on Eq.~(\ref{eq:absr}) for different vorticities $l_{2}$.
The bright structures represent the positions of low light transmission,
while the blue areas correspond to the regions of optical transparency.
As it can be seen, for $l_{2}=1$ an absorption maxima is immersed
in regions of optical transparency (Fig.~\ref{fig:fig3}(a)). For
larger $l_{2}$ numbers, the absorption profile displays a $l_{2}$-fold
symmetry distributed in regions of spatial EIT (Figs.~\ref{fig:fig3}(b)\textendash (e)).
Therefore, one can easily distinguish an unknown vorticity of a vortex
control beam $\Omega_{2}$ just by counting the bright structures
appearing in the absorption profile of the probe field. 

At the core of vortex beam when the intensity of control field $\Omega_{2}$
is zero, the CTL scheme reduces to an M-type atom-light coupling configuration
for which no loss is expected \cite{Luming2005Mtype,Hong2012Mtype}.
Away from the vortex core when $\Omega_{2}\neq0$, the medium is transparent
as long as $|\Omega|^{2}|\beta|^{2}\neq0$ \cite{2017-JPB-Hamedi}
(see also Eqs.~(\ref{eq:linear}) and (\ref{eq:beta^2})). However,
for some specific spatial regions and depending on different vorticities
the parameter $|\Omega|^{2}|\beta|^{2}$ in Eq.~(\ref{eq:beta^2})
may become zero. In this case Eq.~(\ref{eq:linear}) reduces to

\begin{equation}
\rho_{ba}=\frac{\Omega_{p}\left(-|\Omega|^{2}+i\Delta_{p}\left(\Gamma_{e}/2-i\Delta_{p}\right)\right)}{i\left(\Gamma_{e}/2-i\Delta_{p}\right)\zeta+i|\Omega|^{2}\left(\Gamma_{b}/2-i\Delta_{p}\right)+\left(\Gamma_{b}/2-i\Delta_{p}\right)\Delta_{p}\left(\Gamma_{e}/2-i\Delta_{p}\right)}.\label{eq:N-type}
\end{equation}
This is the case when the five-level CTL scheme converts to the conventional
$N$-type atomic structure for which strong absorption is expected
\cite{2017-JPB-Hamedi}. This is the origin of absorption maximums
appearing in absorption profiles illustrated in Figs.~\ref{fig:fig3}(a)\textendash (e).
The $l_{2}$-fold symmetry of absorption profile observed in Fig.~\ref{fig:fig3}
satisfies the predicted cosinusoidal variation of quantum interference
term $Q$ with a periodicity of $l$ given in Eq.~(\ref{eq:QI1}).

\subsubsection{Situation (b): two vortex beams $\Omega_{2}$ and $\Omega_{3}$}

We consider two control fields $\Omega_{2}$ and $\Omega_{3}$ as
vortex beams (Fig.~\ref{fig:schemes}(b)):

\begin{align}
\Omega_{2} & =|\Omega_{2}|\exp(il_{2}\Phi),\label{eq:vortexBeam(b)2}\\
\Omega_{3} & =|\Omega_{3}|\exp(il_{3}\Phi),\label{eq:vortexBeam(b)3}
\end{align}
where $|\Omega_{2}|$ and $|\Omega_{3}|$ are defined by Eq.~\ref{eq:amplitude}
with $j=2,3$. Rabi frequencies of other two nonvortex beams do not
depend on the azimuthal angle $\Phi$:

\begin{align}
\Omega_{1} & =|\Omega_{1}|\,,\label{eq:OtherBeams(b)1}\\
\Omega_{4} & =|\Omega_{4}|\,.\label{eq:OtherBeams(b)4}
\end{align}
In this case, the quantum interference term of Eq.~(\ref{eq:QIT})
takes the form

\begin{equation}
Q=2|\Omega_{1}||\Omega_{2}||\Omega_{3}||\Omega_{4}|\cos\left((l_{2}+l_{3})\Phi\right).\label{eq:QI2}
\end{equation}

\begin{figure}
\includegraphics[width=0.3\columnwidth]{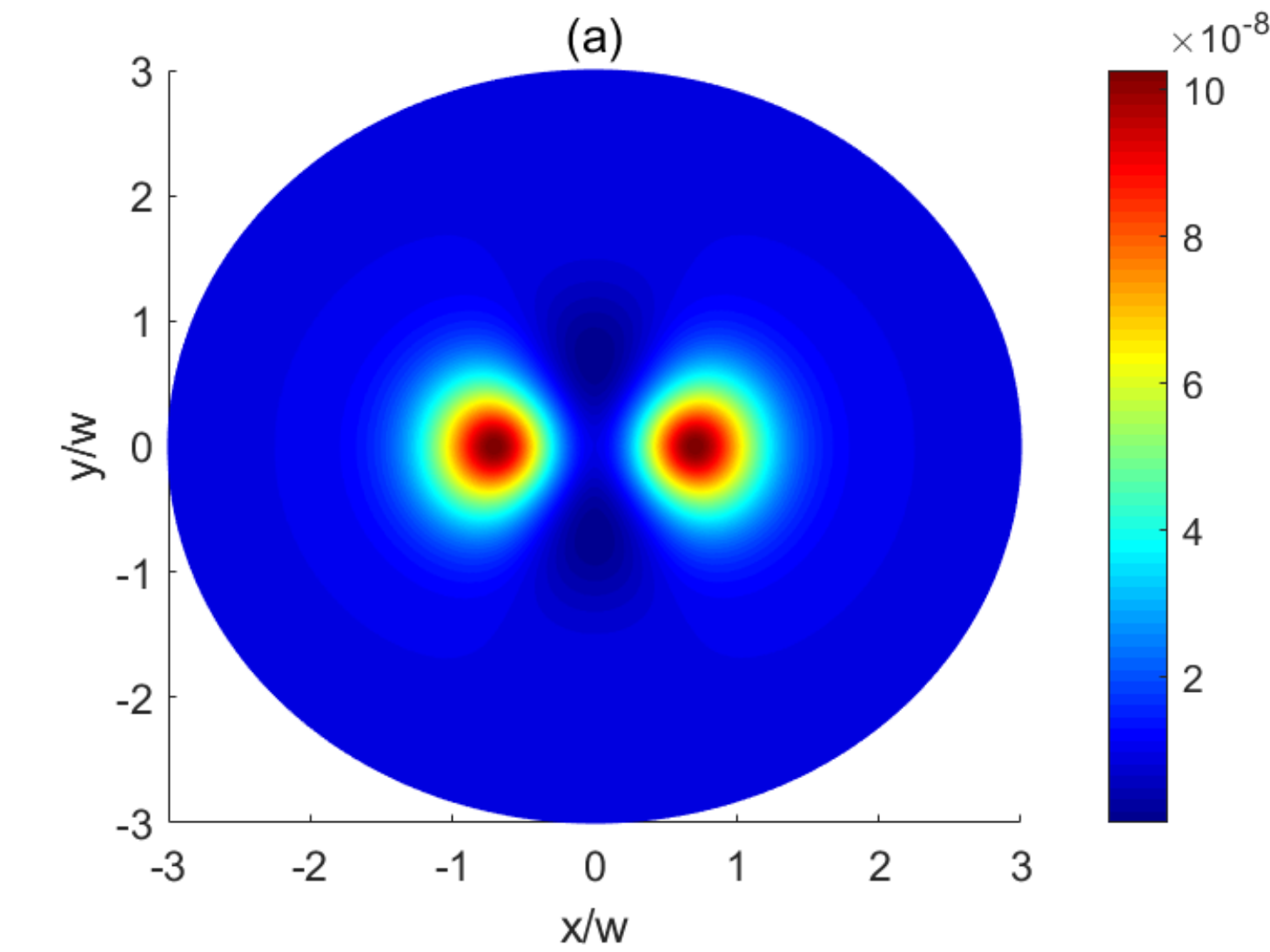} \includegraphics[width=0.3\columnwidth]{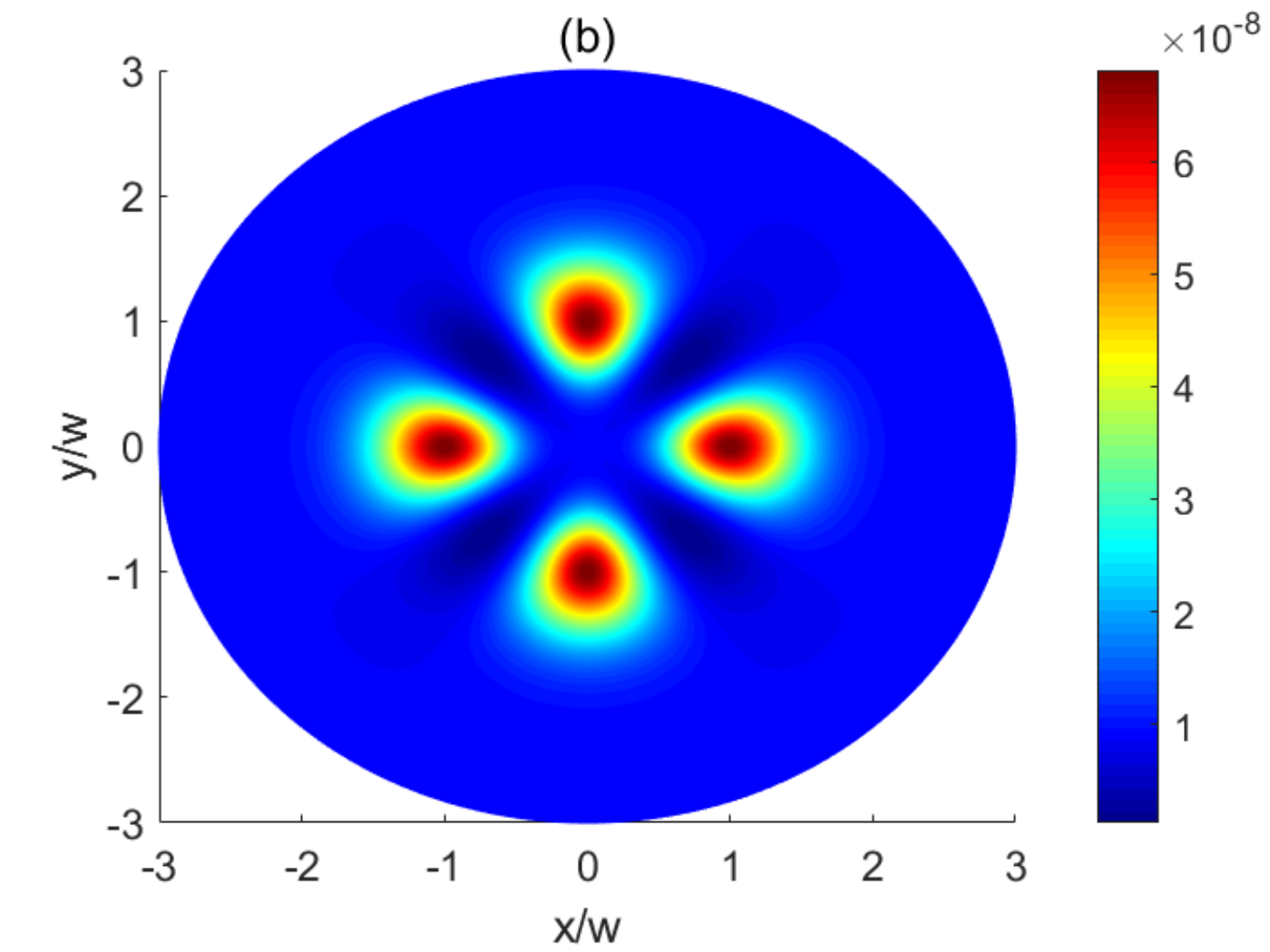}
\includegraphics[width=0.3\columnwidth]{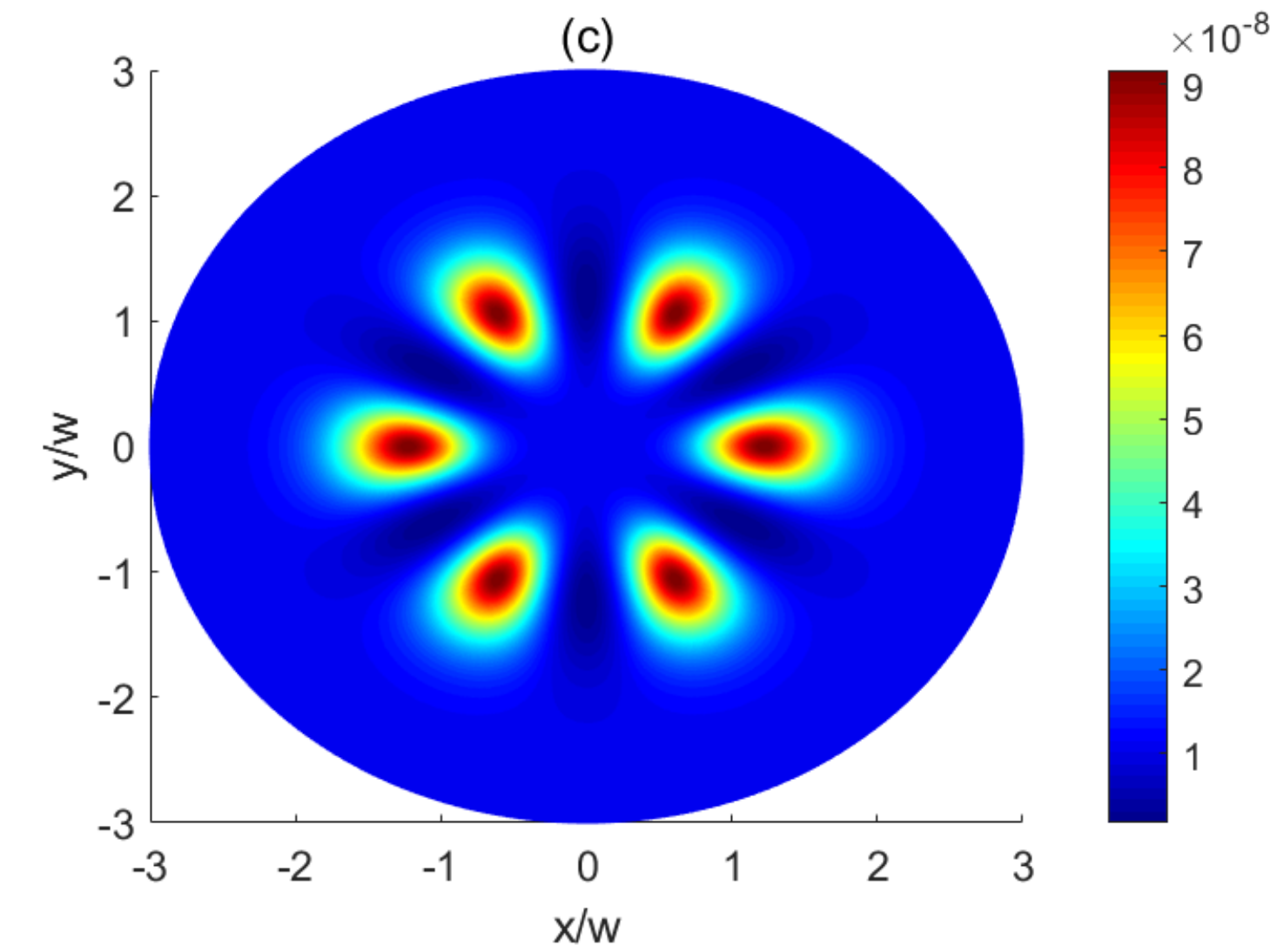}

\includegraphics[width=0.3\columnwidth]{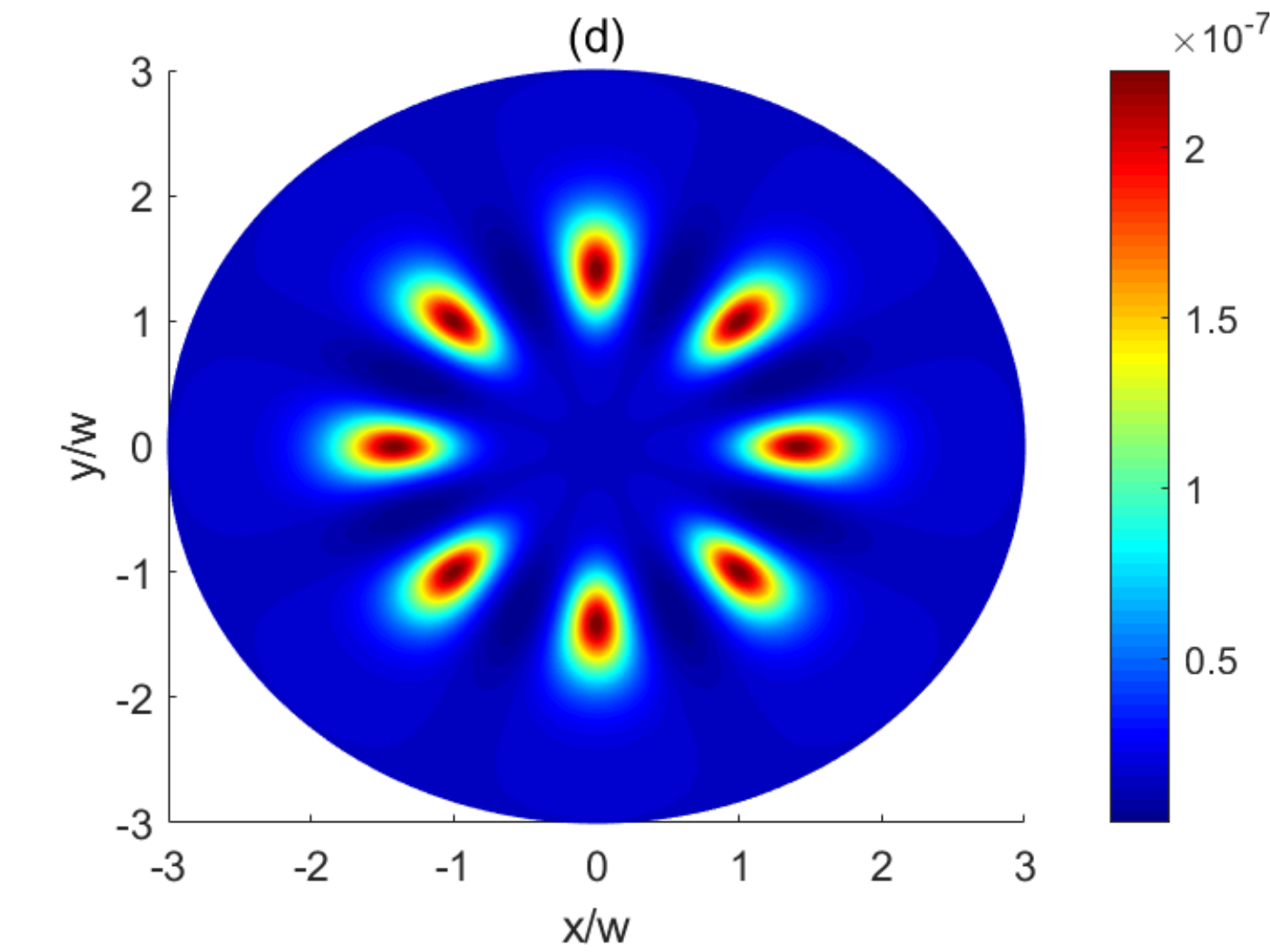} \includegraphics[width=0.3\columnwidth]{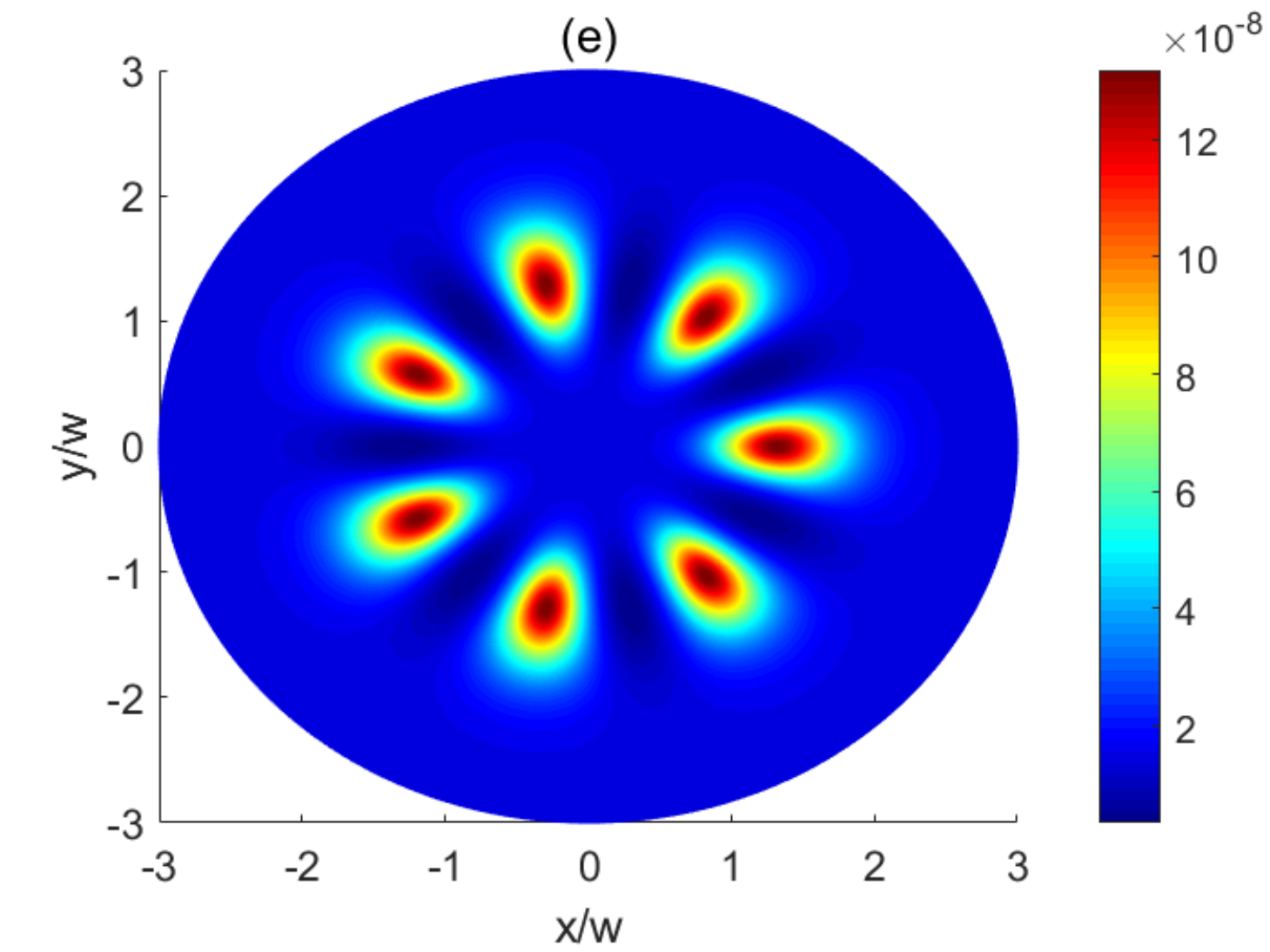}
\caption{Spatially dependent absorption profile of the probe beam in arbitrary
units when two control fields $\Omega_{2}$ and $\Omega_{3}$ have
optical vortices. The vorticities are $l=1$ (a), $l=2$ (b), $l=3$
(c), $l=4$ (d) and $l_{2}=3$, $l_{3}=4$ (e). The remaining parameters
are $|\Omega_{1}|=0.6\Gamma,$ $\epsilon_{2}=0.7\Gamma,$ $\epsilon_{3}=0.3\Gamma,$
$|\Omega_{4}|=0.5\Gamma,$ and the other parameters are the same as
Fig.~\ref{fig:fig3}.}
\label{fig:fig4}
\end{figure}

In Fig.~\ref{fig:fig4} we show the absorption profile when $\Omega_{2}$
and $\Omega_{3}$ are structured light. Let us first consider that
the vortices are equal: $l_{2}=l_{3}\equiv l$ . In this case, Eq.~(\ref{eq:QI2})
reduces to

\begin{equation}
Q=2|\Omega_{1}||\Omega_{2}||\Omega_{3}||\Omega_{4}|\cos(2l\Phi),\label{eq:QI2-1}
\end{equation}
and the spatial profile of probe absorption displays a $2l$-fold
symmetry (Figs.~\ref{fig:fig4} (a)\textendash (d)) satisfying the
$2l$ cosinusoidal behavior of quantum interference term featured
in Eq.~(\ref{eq:QI2-1}). The medium is transparent at the core of
vortex beams as the five-level CTL scheme is now equivalent to a three-level
$\Lambda$-type system which is decoupled from the two level system
involving the states $|e\rangle$ and $|d\rangle$ \cite{2017-JPB-Hamedi}.

When the vortices are different $l_{2}\neq l_{3}$, the number of
absorption peaks becomes $|l_{2}+l_{3}|$, and the symmetry of absorption
profile obeys from  the predicted cosinusoidal behavior of the parameter
$Q$ with a periodicity of $|l_{2}+l_{3}|$, given in Eq.~(\ref{eq:QI2}).
For instance, for the case of different vortices $l_{2}=3$ and $l_{3}=4$
the number of symmetrical absorption peaks becomes seven as illustrated
in Fig.~\ref{fig:fig4}(e).

\subsubsection{Situation (c): two vortex beams $\Omega_{3}$ and $\Omega_{4}$}

In the situation, shown in Fig.~\ref{fig:schemes}(c), the control
fields with Rabi frequencies $\Omega_{3}$ and $\Omega_{4}$ are assumed
to carry optical vortices

\begin{align}
\Omega_{3} & =|\Omega_{3}|\exp(il_{3}\Phi),\label{eq:vortexBeam(c)3}\\
\Omega_{4} & =|\Omega_{4}|\exp(il_{4}\Phi),\label{eq:vortexBeam(c)4}
\end{align}
where $|\Omega_{3}|$ and $|\Omega_{4}|$ are defined by Eq.~\ref{eq:amplitude}
with $j=3,4$, yet other control beams are nonvortex beams,

\begin{align}
\Omega_{1} & =|\Omega_{1}|\,,\label{eq:OtherBeams(c)1}\\
\Omega_{2} & =|\Omega_{2}|\,.\label{eq:OtherBeams(c)2}
\end{align}
Equation.~(\ref{eq:QIT}) then becomes

\begin{equation}
Q=2|\Omega_{1}||\Omega_{2}||\Omega_{3}||\Omega_{4}|\cos\left((l_{3}-l_{4})\Phi\right).\label{eq:QI3}
\end{equation}

\begin{figure}
\includegraphics[width=0.3\columnwidth]{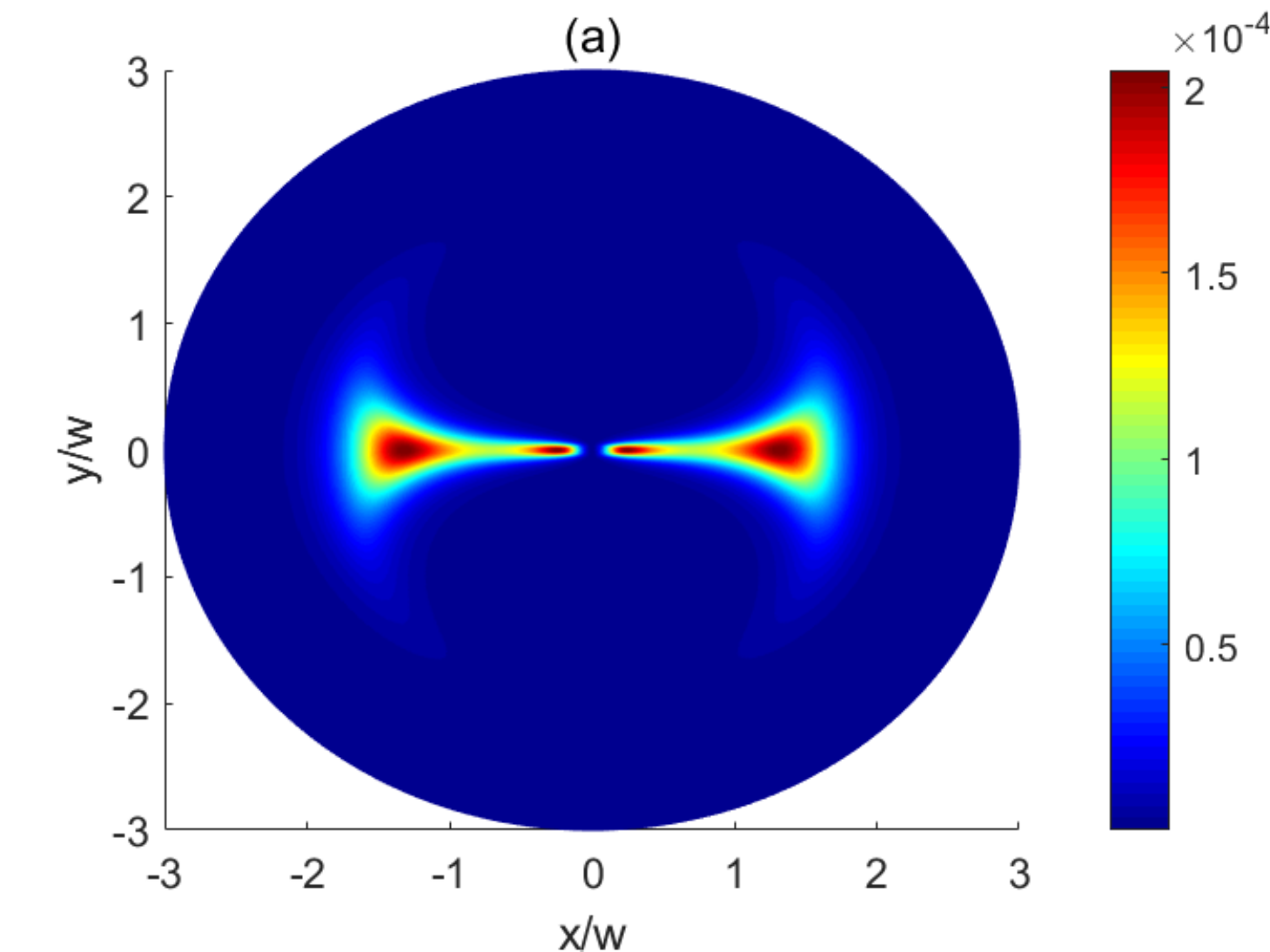} \includegraphics[width=0.3\columnwidth]{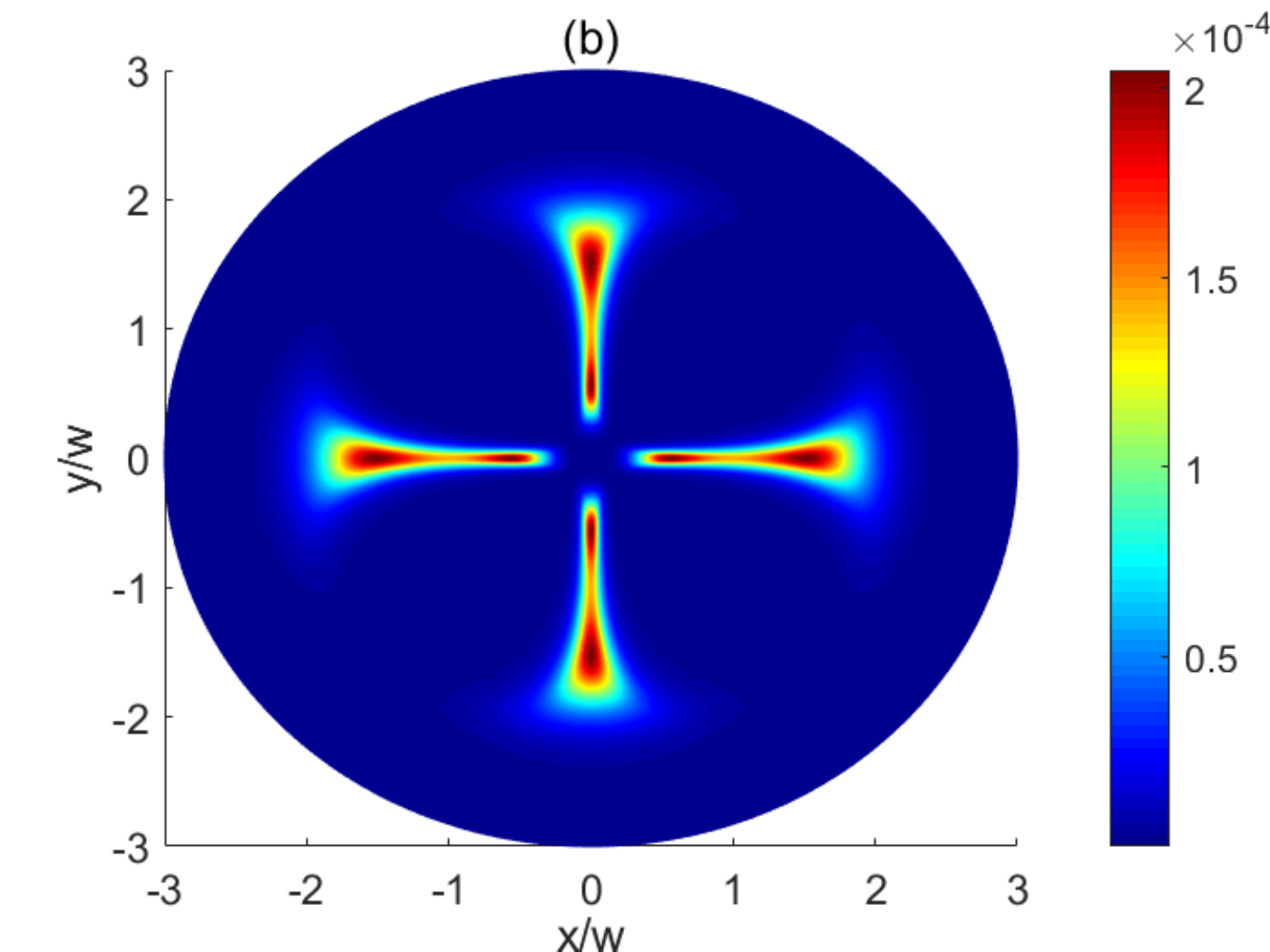}
\includegraphics[width=0.3\columnwidth]{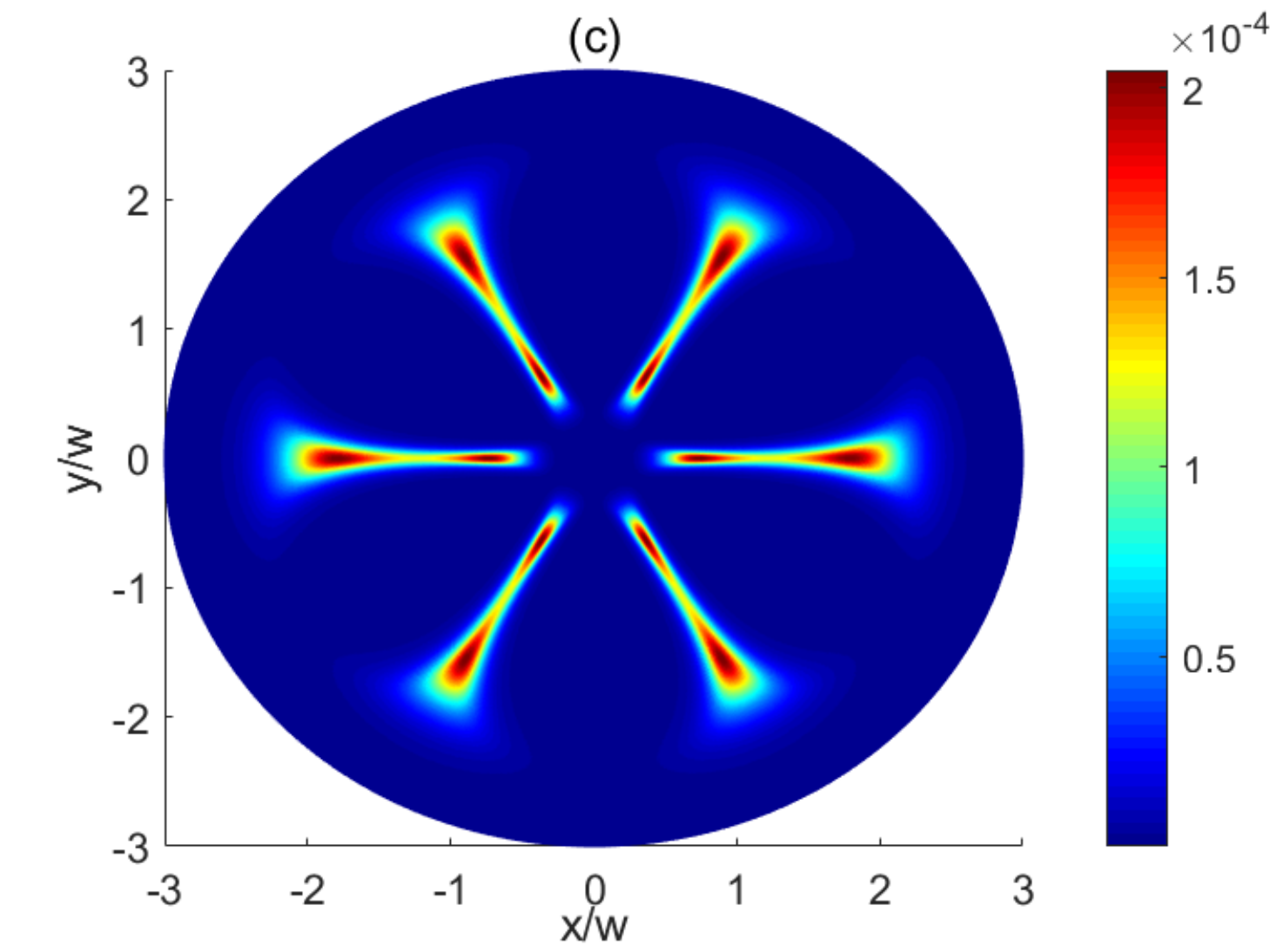}

\includegraphics[width=0.3\columnwidth]{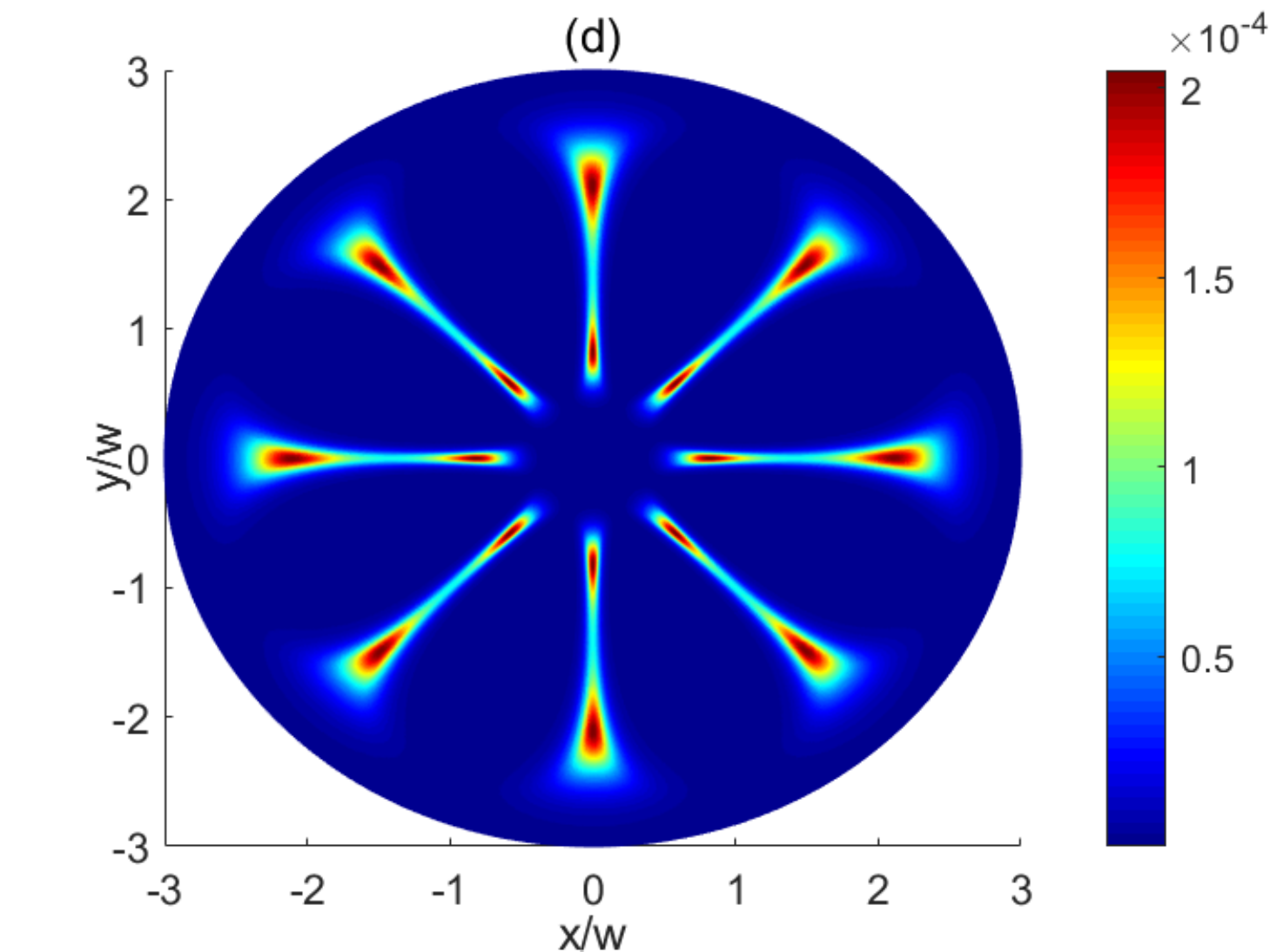} \includegraphics[width=0.3\columnwidth]{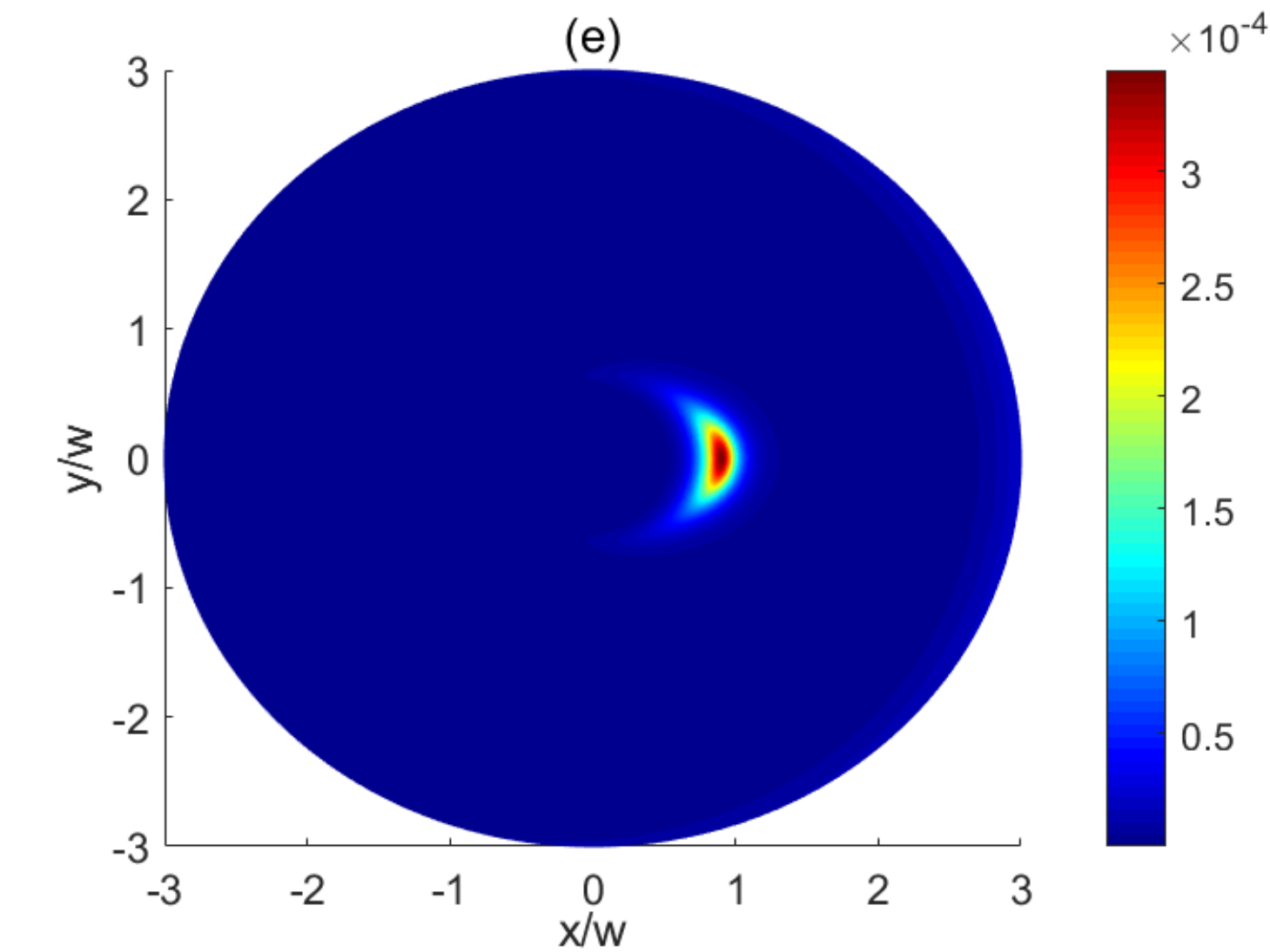}
\caption{Spatially dependent absorption profile of the probe beam in arbitrary
units when two control fields $\Omega_{3}$ and $\Omega_{4}$ have
optical vortices. The vorticities are $l=1$ (a), $l=2$ (b), $l=3$
(c), $l=4$ (d) and $l_{3}=5$, $l_{4}=6$ (e). The selected parameters
are $|\Omega_{1}|=0.6\Gamma,$ $|\Omega_{2}|=0.7\Gamma,$ $\epsilon_{3}=0.3\Gamma,$
$\epsilon_{4}=0.5\Gamma,$ and the other parameters are the same as
Fig.~\ref{fig:fig3}. }
\label{fig:fig5}
\end{figure}

When $r\rightarrow0$, the system simplifies to a tripod atom-light
coupling structure, hence, we expect optical transparency at the core
of vortices \cite{2002-JOB-Paspalakis} (see Fig.~\ref{fig:fig5}).
For the opposite helicity optical vortices $l_{3}=-l_{4}\equiv l$,
the quantum interference term $Q$ is the same as Eq.~(\ref{eq:QI2-1}),
yielding a $2l$-fold symmetry of structured transparency profile,
as illustrated in Figs.~\ref{fig:fig5} (a)\textendash (d). When
$l_{3}\neq l_{4}$, Eq.~(\ref{eq:QI3}) necessitates a $|l_{3}-l_{4}|$
symmetry of the absorption profile (Fig.~\ref{fig:fig5} (e)). 

\subsubsection{Situation (d): two vortex beams $\Omega_{2}$ and $\Omega_{4}$}

\begin{figure}
\includegraphics[width=0.3\columnwidth]{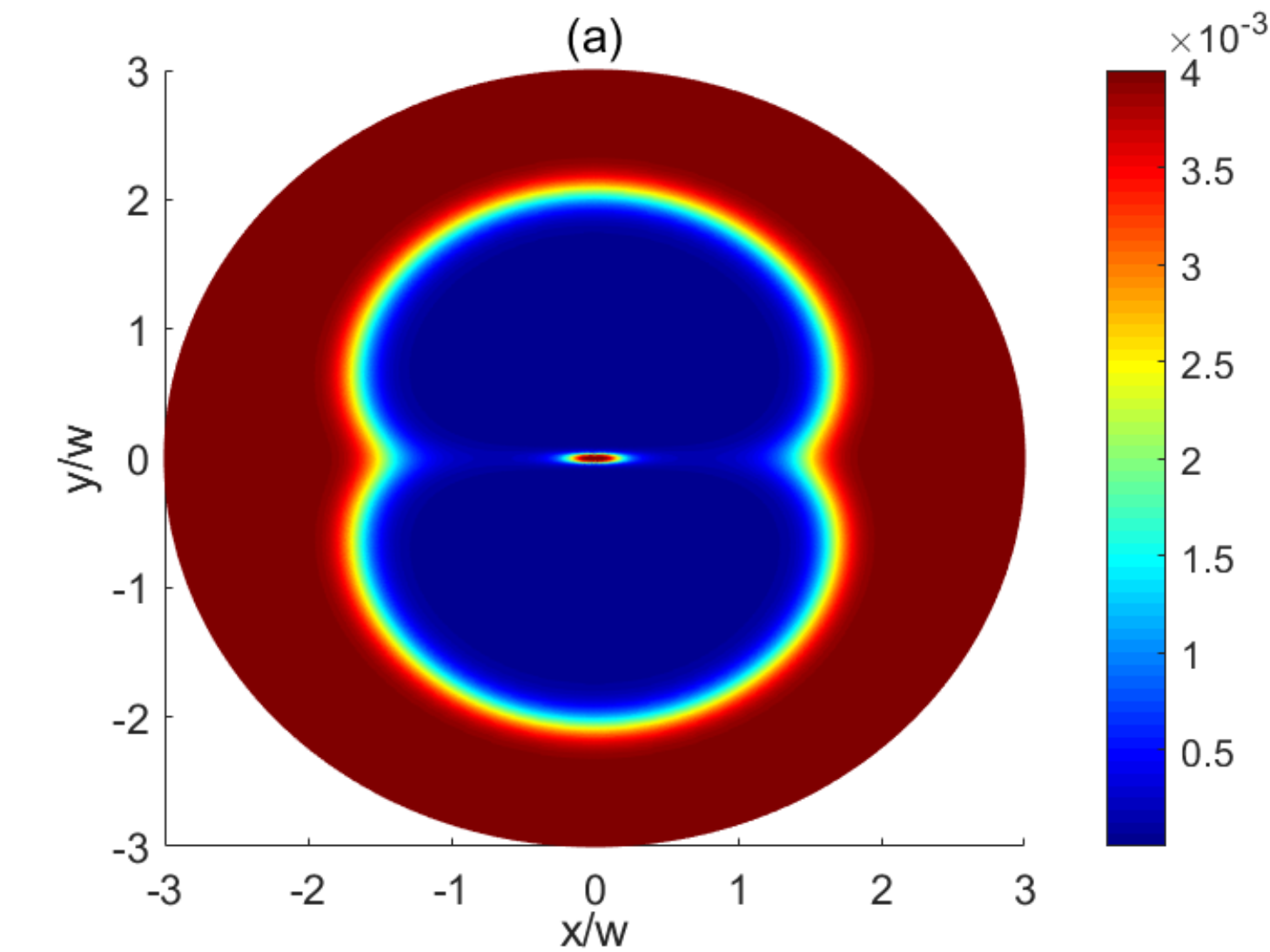} \includegraphics[width=0.3\columnwidth]{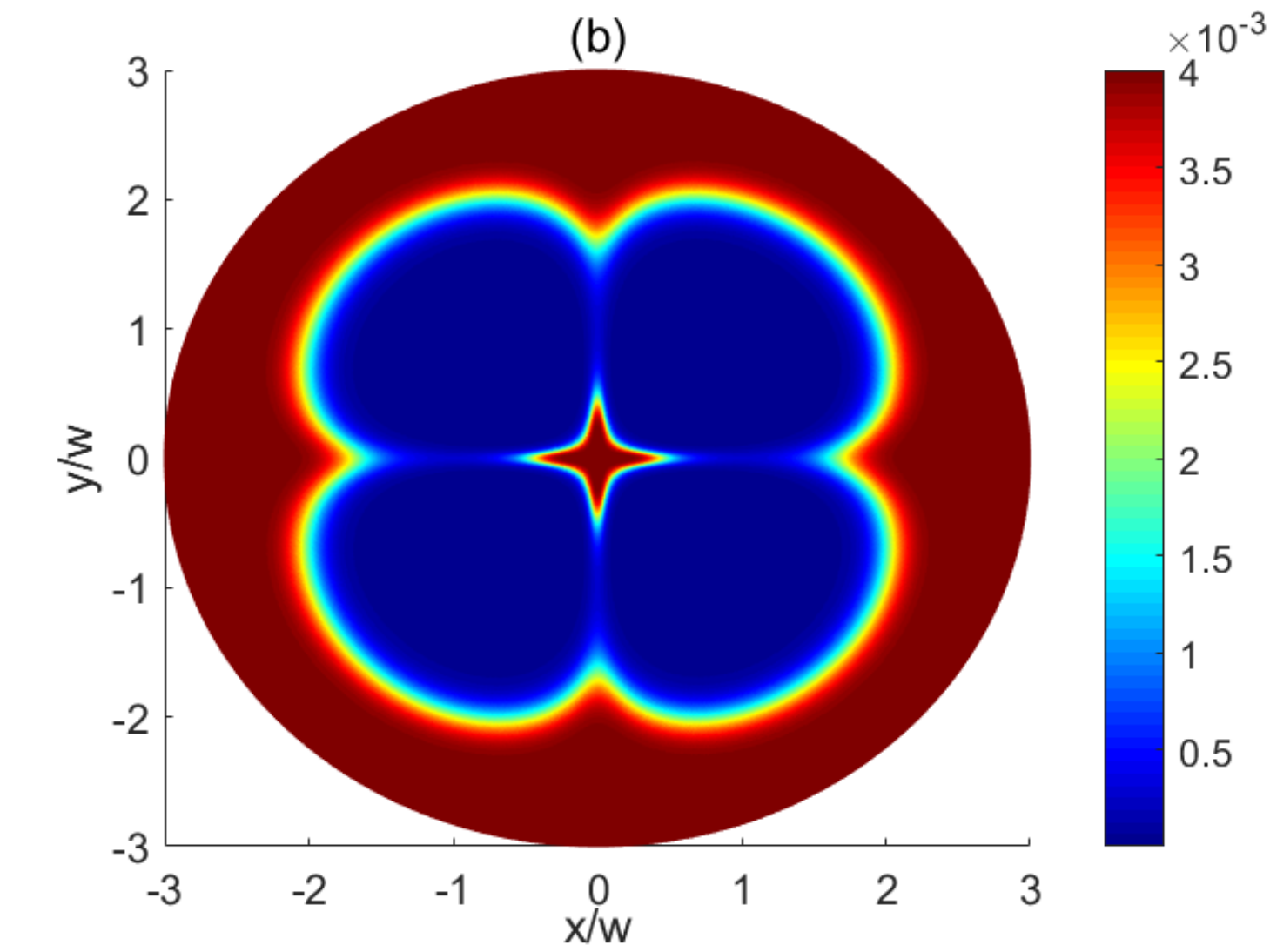}
\includegraphics[width=0.3\columnwidth]{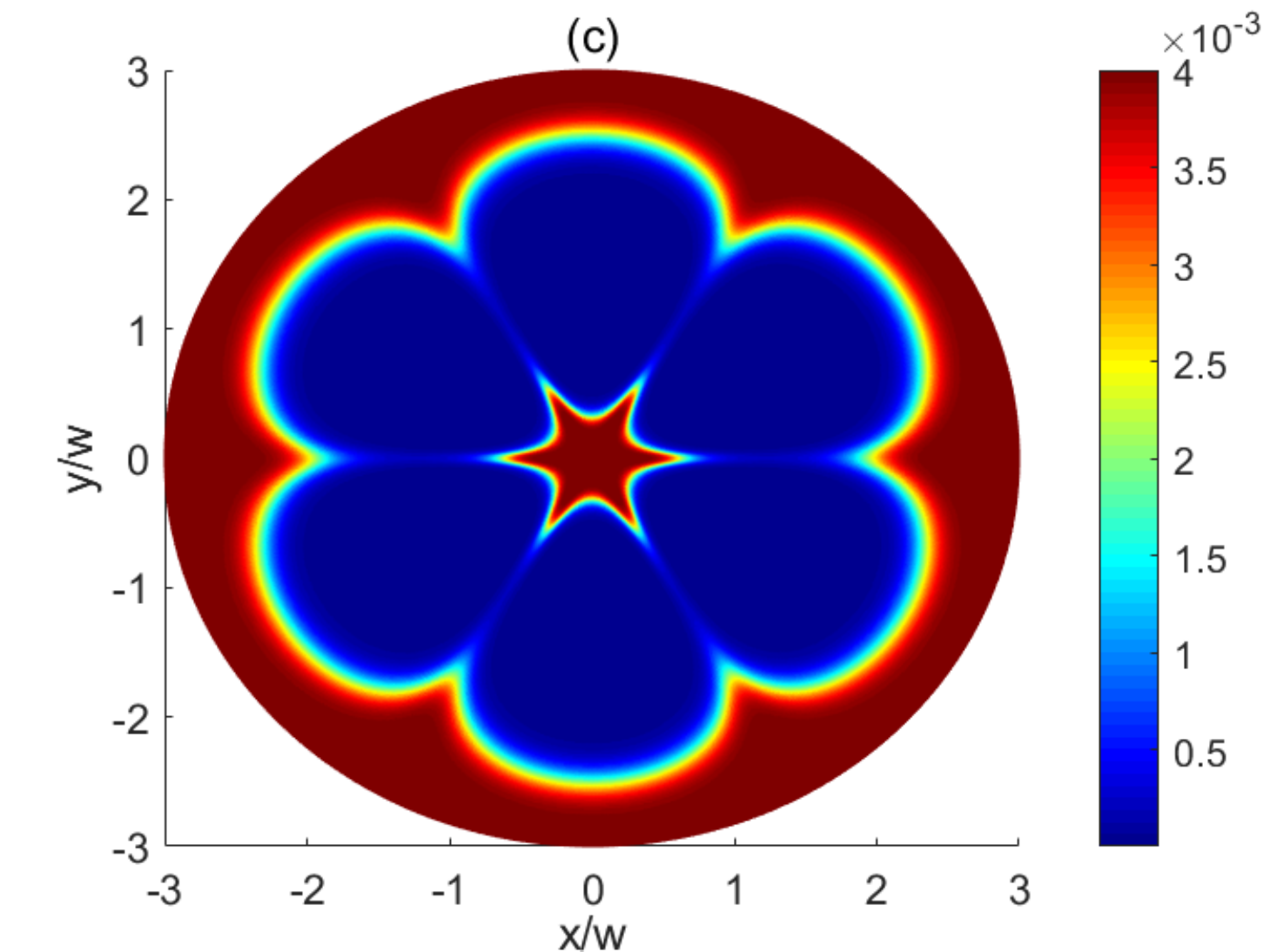}

\includegraphics[width=0.3\columnwidth]{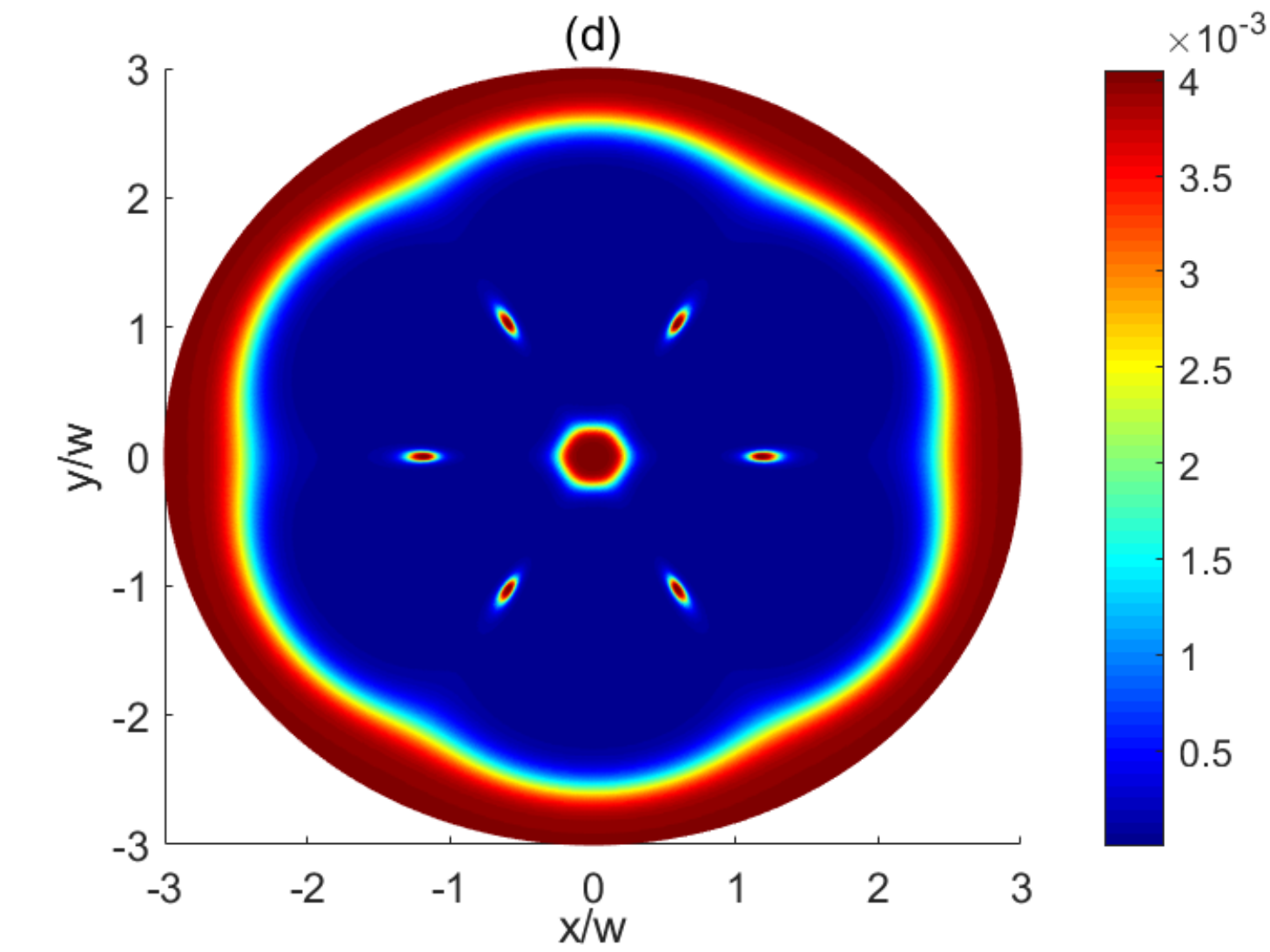} \includegraphics[width=0.3\columnwidth]{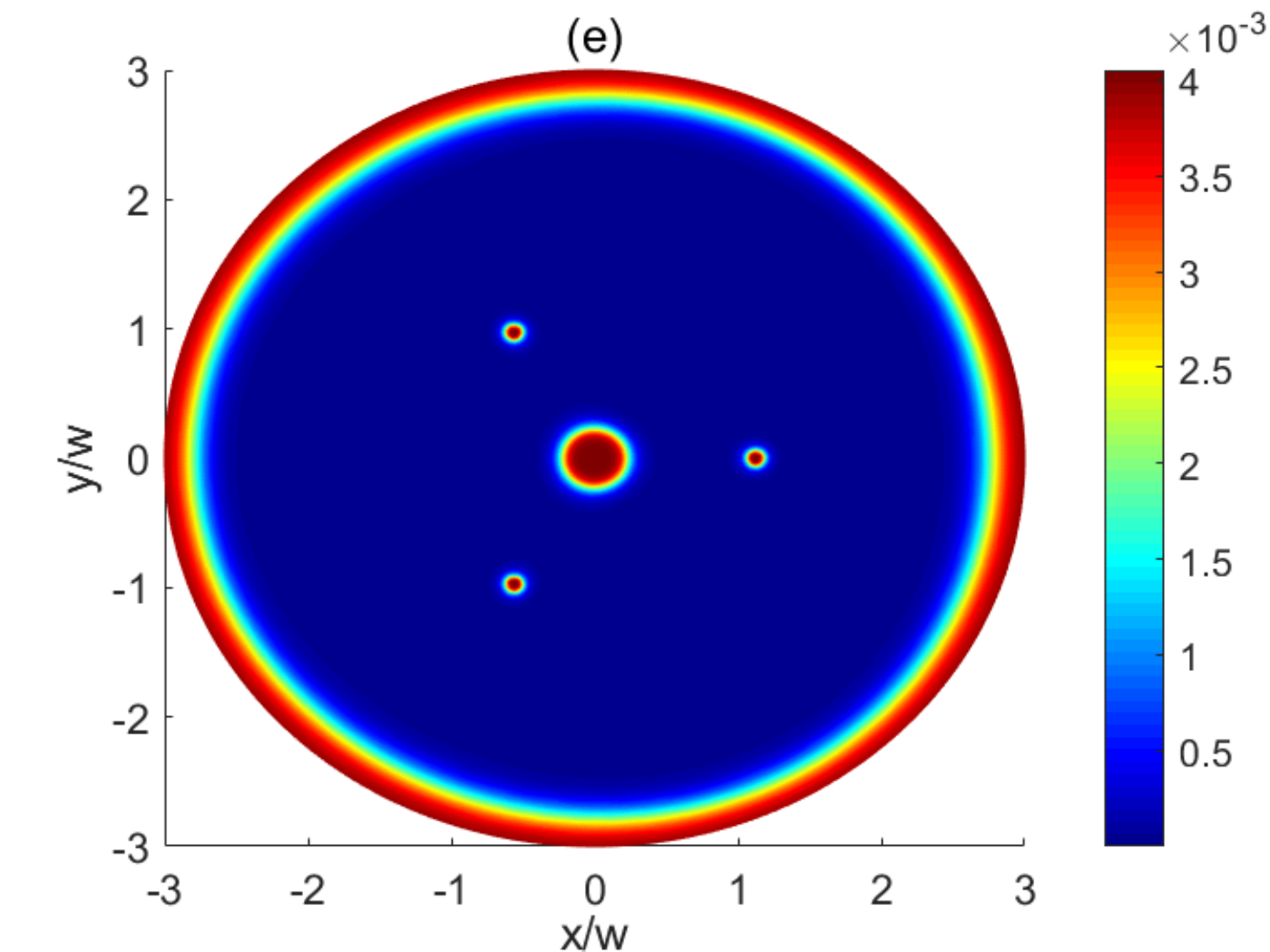}
\caption{Spatially dependent absorption profile of the probe beam in arbitrary
units when two control fields $\Omega_{2}$ and $\Omega_{4}$ have
optical vortices. The vorticities are $l=1$ (a), $l=2$ (b),$l=3$
(c), $l_{2}=4$, $l_{4}=-2$ (d) and $l_{2}=5$, $l_{4}=2$ (e). The
selected parameters are $|\Omega_{1}|=0.6\Gamma,$ $\epsilon_{2}=0.7\Gamma,$
$|\Omega_{3}|=0.3\Gamma,$ $\epsilon_{4}=0.5\Gamma,$ and the other
parameters are the same as Fig.~\ref{fig:fig3}. }
\label{fig:fig6}
\end{figure}

Now we assume that the two control fields $\Omega_{2}$ and $\Omega_{4}$
are structured light (Fig.~\ref{fig:schemes}(e)),

\begin{align}
\Omega_{2} & =|\Omega_{2}|\exp(il_{2}\Phi),\label{eq:vortexBeam(e)2}\\
\Omega_{4} & =|\Omega_{4}|\exp(il_{4}\Phi),\label{eq:vortexBeam(e)4}
\end{align}
where $|\Omega_{2}|$ and $|\Omega_{4}|$ are defined by Eq.~\ref{eq:amplitude}
with $j=2,4$, and

\begin{align}
\Omega_{1} & =|\Omega_{1}|\,,\label{eq:OtherBeams(e)1}\\
\Omega_{3} & =|\Omega_{3}|\,.\label{eq:OtherBeams(e)3}
\end{align}
The quantum interference of Eq.~(\ref{eq:QIT}) becomes

\begin{equation}
Q=2|\Omega_{1}||\Omega_{2}||\Omega_{3}||\Omega_{4}|\cos\left((l_{2}-l_{4})\Phi\right).\label{eq:QI5}
\end{equation}
For the opposite helicity vortex beams $l_{2}=-l_{4}=l$, the quantum
interference term changes to Eq.~(\ref{eq:QI2-1}) and the absorption
profile demonstrates again a $2l$- fold symmetry (Figs.~\ref{fig:fig6}(a)\textendash (c)). 

When the vortices are not the same $l_{2}\neq l_{4}$, the symmetry
of spatial absorption profile follows from Eq.~\ref{eq:QI5}, for
example for $l_{2}=4$ and $l_{4}=-2$ (Fig.~\ref{fig:fig6}(d))
and $l_{2}=5$ and $l_{4}=2$ (Fig.~\ref{fig:fig6}(e)), $6$-fold
and $3$-fold symmetries are observed. Note that the core of absorption
image now corresponds to regions of low light transmission as for
this situation the medium becomes an absorptive $N$-type medium \cite{Sheng2011Ntype,Yang2015Ntype}
at the core of control vortex beams. 

\subsubsection{Situation (e): three vortex beams $\Omega_{2}$, $\Omega_{3}$ and
$\Omega_{4}$}

\begin{figure}
\includegraphics[width=0.3\columnwidth]{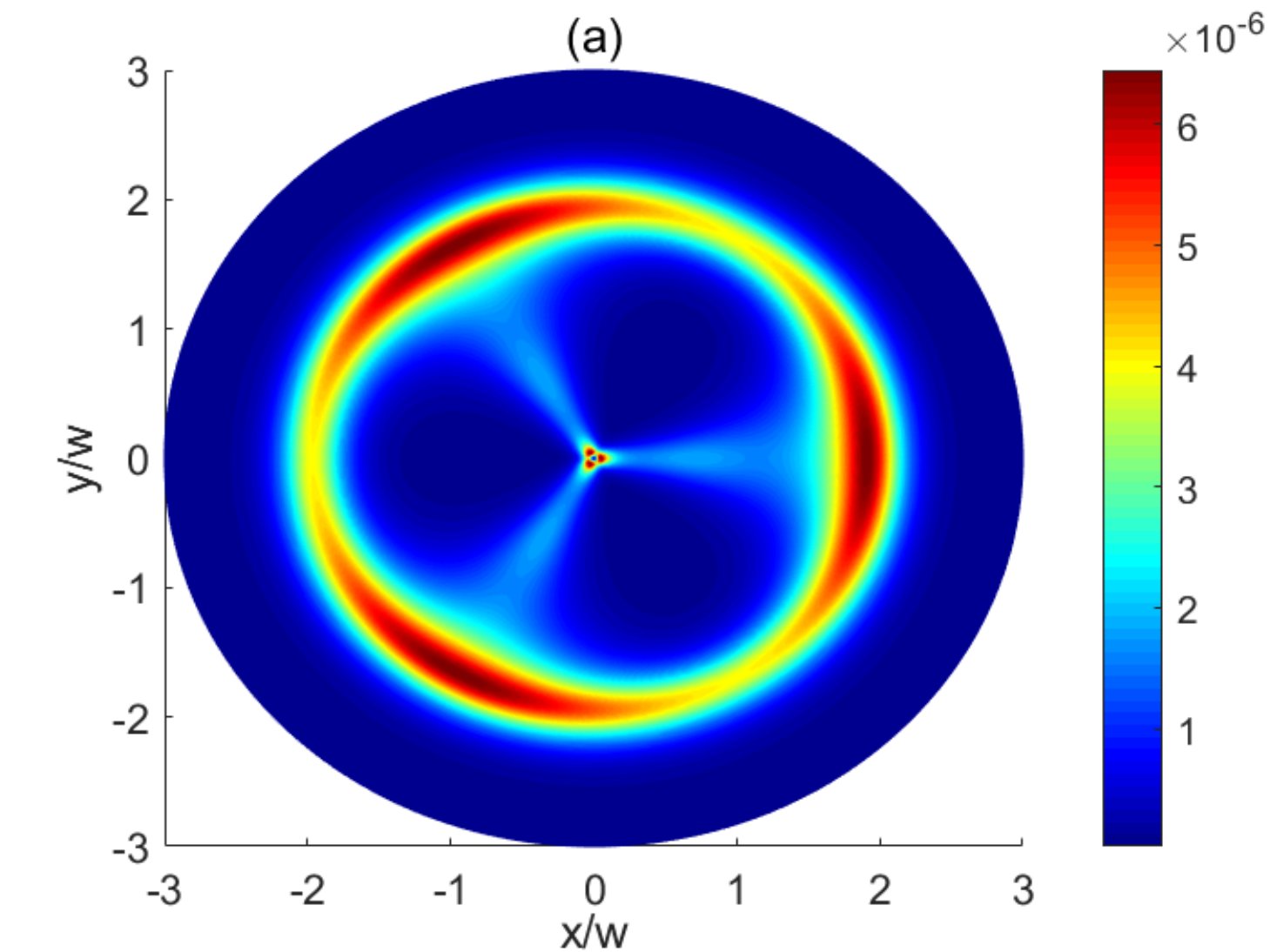} \includegraphics[width=0.3\columnwidth]{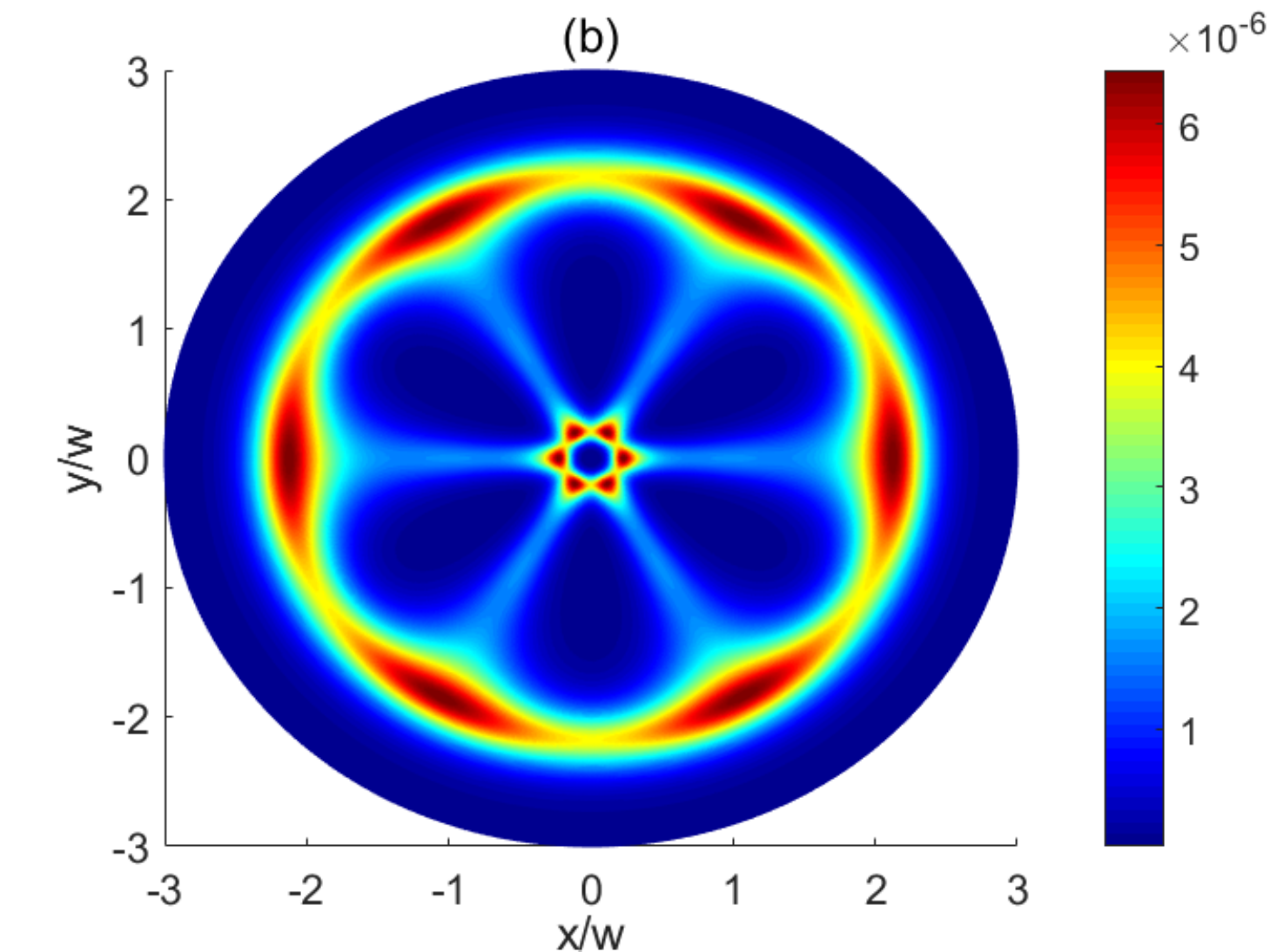}
\includegraphics[width=0.3\columnwidth]{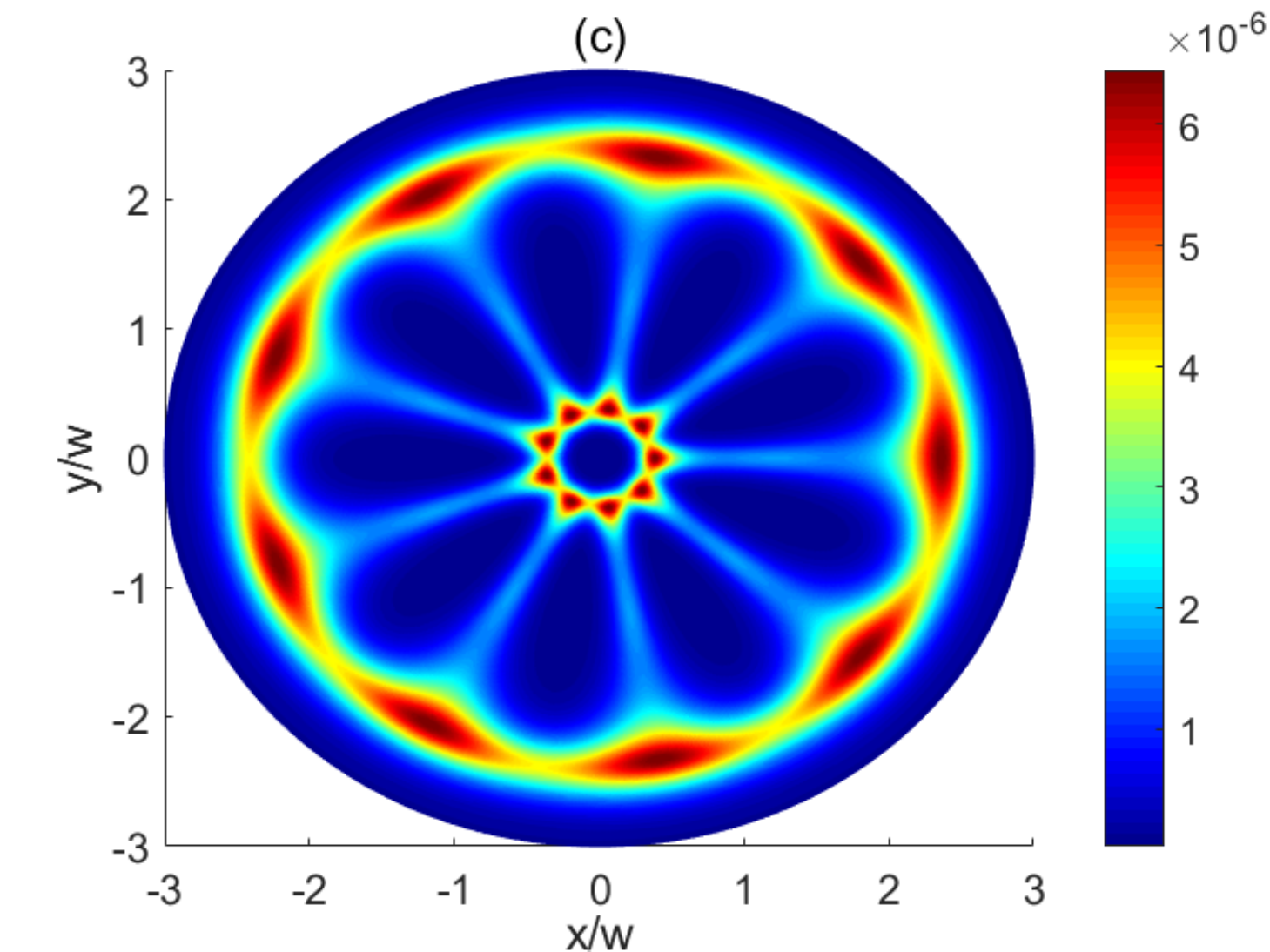}

\includegraphics[width=0.3\columnwidth]{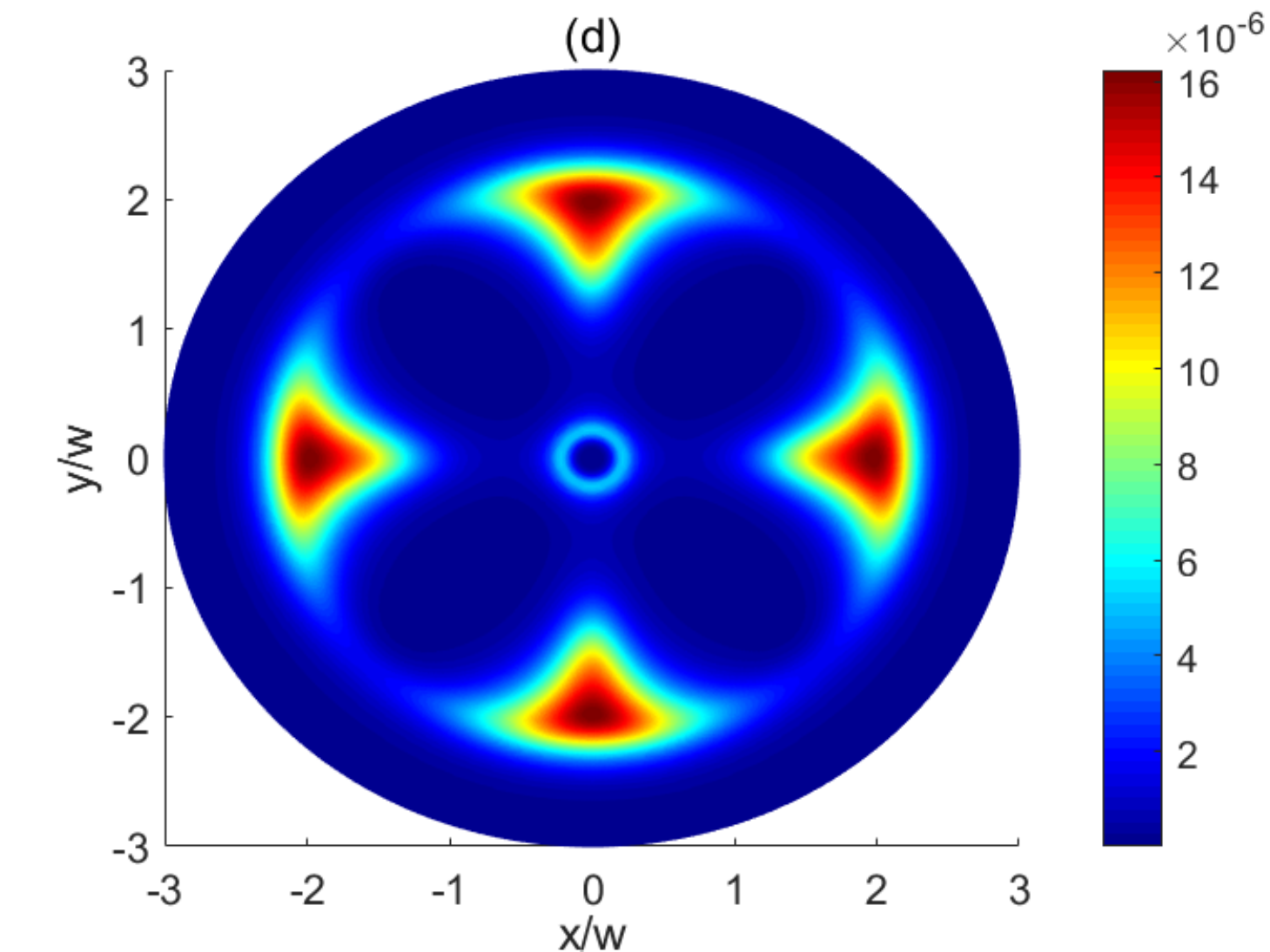} \includegraphics[width=0.3\columnwidth]{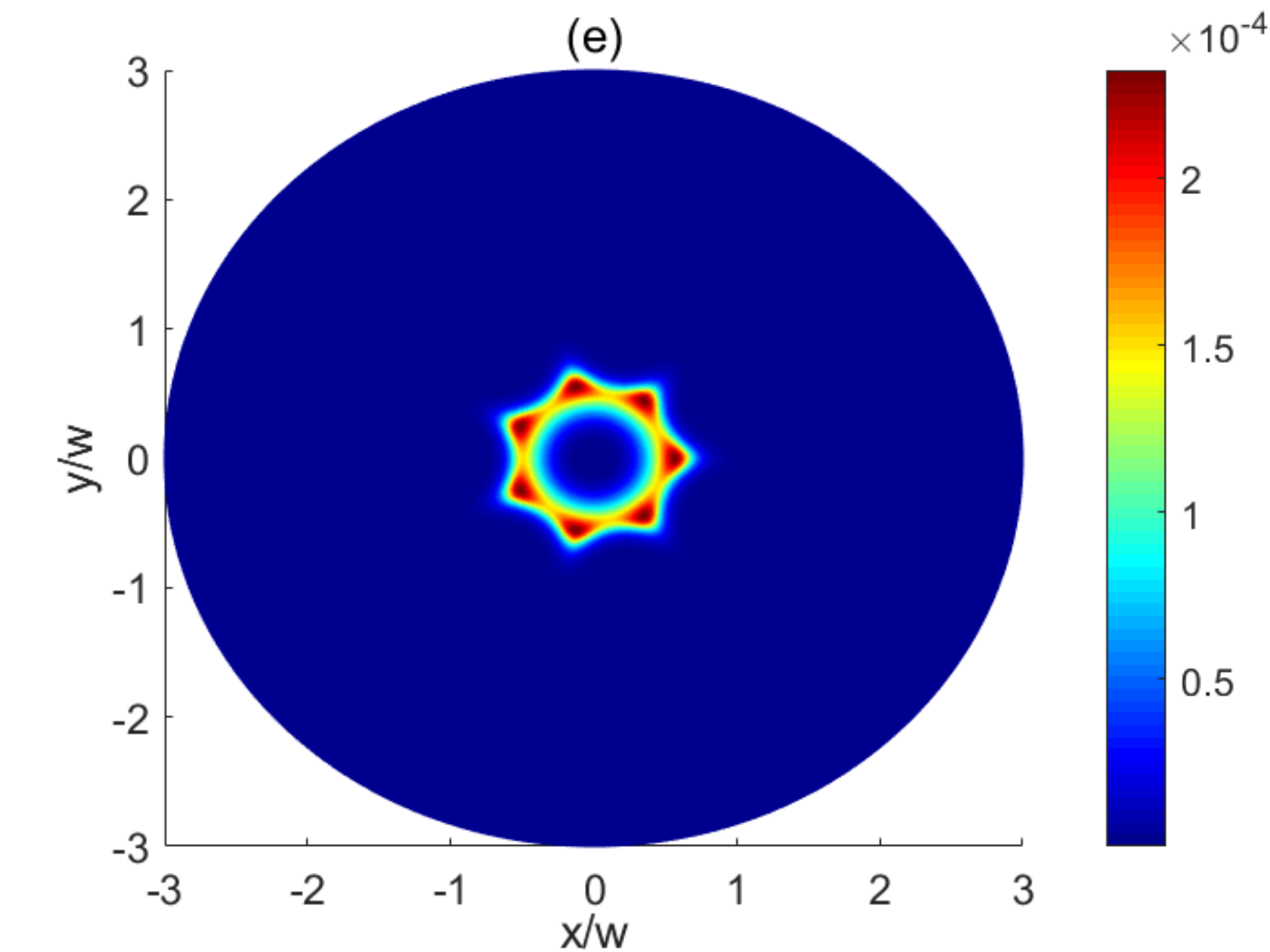}
\caption{Spatially dependent absorption profile of the probe beam in arbitrary
units when three control fields $\Omega_{2}$, $\Omega_{3}$ and $\Omega_{4}$
have optical vortices. The vorticities are $l=1$ (a), $l=2$ (b),
$l=3$ (c), $l_{2}=4$, $l_{3}=l_{4}=-2$ (d) and $l_{2}=4$, $l_{3}=-2$,
$l_{4}=-5$ (e). The selected parameters are $|\Omega_{1}|=0.6\Gamma,$
$\epsilon_{2}=0.7\Gamma,$ $\epsilon_{3}=0.3\Gamma,$ $\epsilon_{4}=0.5\Gamma,$
and the other parameters are the same as Fig.~\ref{fig:fig3}. }
\label{fig:fig7}
\end{figure}

We consider three control fields $\Omega_{2}$, $\Omega_{3}$ and
$\Omega_{4}$ as vortex beams (Fig.~\ref{fig:schemes}(d)),

\begin{align}
\Omega_{2} & =|\Omega_{2}|\exp(il_{2}\Phi),\label{eq:vortexBeam(d)2}\\
\Omega_{3} & =|\Omega_{3}|\exp(il_{3}\Phi),\label{eq:vortexBeam(d)3}\\
\Omega_{4} & =|\Omega_{4}|\exp(il_{4}\Phi),\label{eq:vortexBeam(d)4}
\end{align}
with $|\Omega_{2}|$, $|\Omega_{3}|$and $|\Omega_{4}|$ defined by
Eq.~\ref{eq:amplitude} and $j=2,3,4$. The nonvortex beam has the
Rabi frequency

\begin{equation}
\Omega_{1}=|\Omega_{1}|\,,\label{eq:OtherBeams(d)1}
\end{equation}
In this case, the quantum interference term of Eq.~(\ref{eq:QIT})
reads

\begin{equation}
Q=2|\Omega_{1}||\Omega_{2}||\Omega_{3}||\Omega_{4}|\cos\left((l_{2}+l_{3}-l_{4})\Phi\right).\label{eq:QI4}
\end{equation}
When two control beams $\Omega_{2}$ and $\Omega_{3}$ have the same
vortices while they are of opposite helicity with the vortex beam
$\Omega_{4}$ ($l_{2}=l_{3}=-l_{4}\equiv l$), the quantum interference
term becomes

\begin{equation}
Q=2|\Omega_{1}||\Omega_{2}||\Omega_{3}||\Omega_{4}|\cos\left(3l\Phi\right).\label{eq:QI4-1}
\end{equation}
Obviously, the $3l$ cosinusoidal behavior of $Q$ given in Eq.~(\ref{eq:QI4-1})
leads to a $3l$-fold symmetry of spatial distribution of absorption
profile, as depicted in Figs.~\ref{fig:fig7}(a)\textendash (c).
The symmetry of patterns changes to $l_{2}+l_{3}-l_{4}$-fold for
$l_{2}\neq l_{3}\neq l_{4}$(Figs.~\ref{fig:fig7}(d)\textendash (e)).
Note that the medium is now equivalent to a three-level $\Lambda$
scheme at the core of control vortex beams,resulting in an optical
transparency for $r\rightarrow0$. 

\subsubsection{Situation (f): all control fields are vortex beams}

Finally, let us consider a situation when all control fields are assumed
to be vortex beams (Fig.~\ref{fig:schemes}(f))

\begin{align}
\Omega_{1} & =|\Omega_{1}|\exp(il_{1}\Phi),\label{eq:vortexBeam(f)1}\\
\Omega_{2} & =|\Omega_{2}|\exp(il_{2}\Phi),\label{eq:vortexBeam(f)2}\\
\Omega_{3} & =|\Omega_{3}|\exp(il_{3}\Phi),\label{eq:vortexBeam(f)3}\\
\Omega_{4} & =|\Omega_{4}|\exp(il_{4}\Phi),\label{eq:vortexBeam(f)4}
\end{align}
with$|\Omega_{1}|$, $|\Omega_{2}|$, $|\Omega_{3}|$ and $|\Omega_{4}|$
defined by Eq.~\ref{eq:amplitude} where $j=1,2,3,4$. For this situation
the system reduces to a two-level absorptive medium at the core of
optical vortices, therefore we expect to observe low light transmission
in the absorption image of probe field when $r\rightarrow0$. The
quantum interference term Eq.~(\ref{eq:QIT}) changes to

\begin{equation}
Q=2|\Omega_{1}||\Omega_{2}||\Omega_{3}||\Omega_{4}|\cos\left((l_{2}+l_{3}-l_{1}-l_{4})\Phi\right).\label{eq:QI6}
\end{equation}

\begin{figure}
\includegraphics[width=0.3\columnwidth]{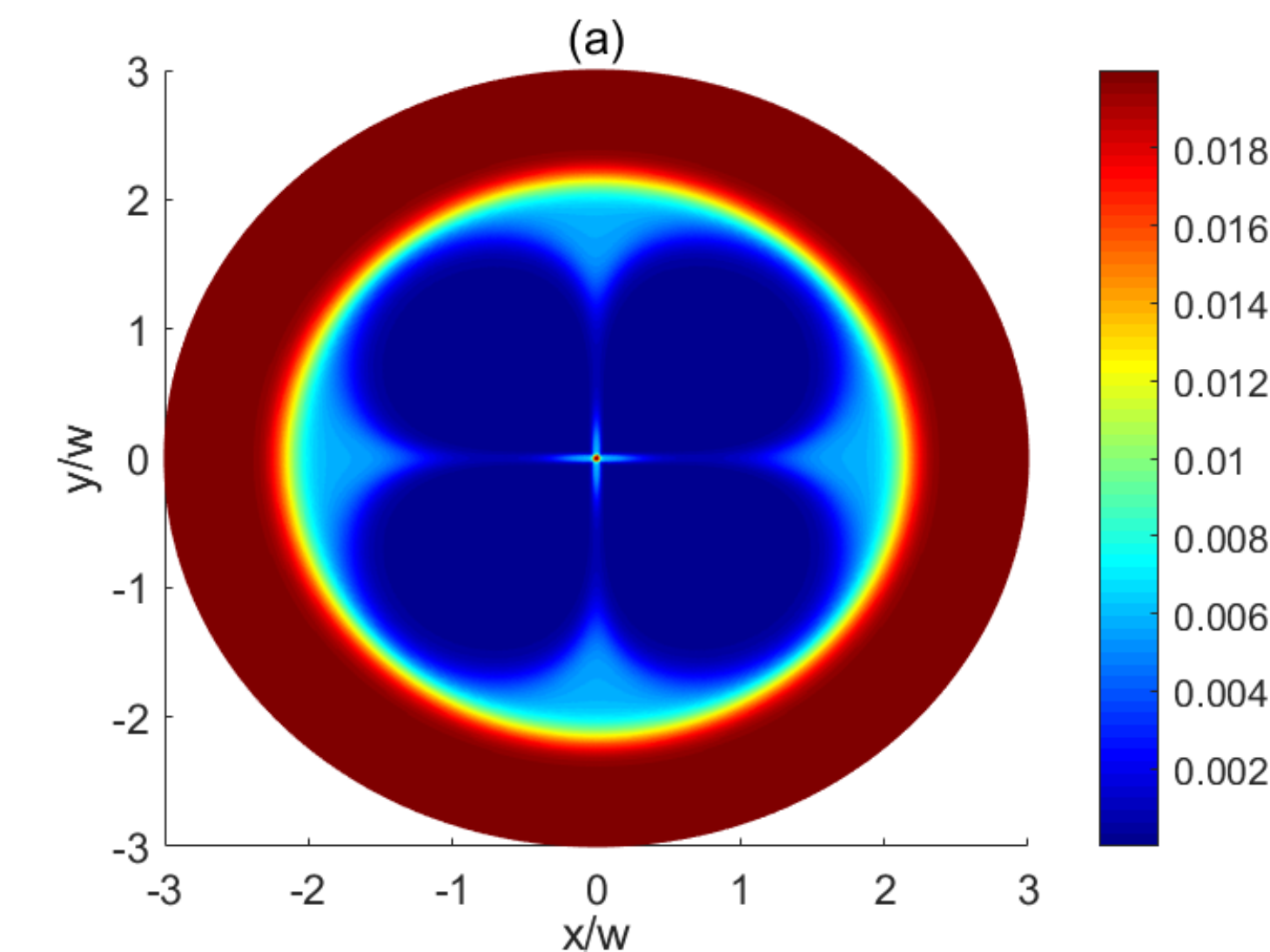} \includegraphics[width=0.3\columnwidth]{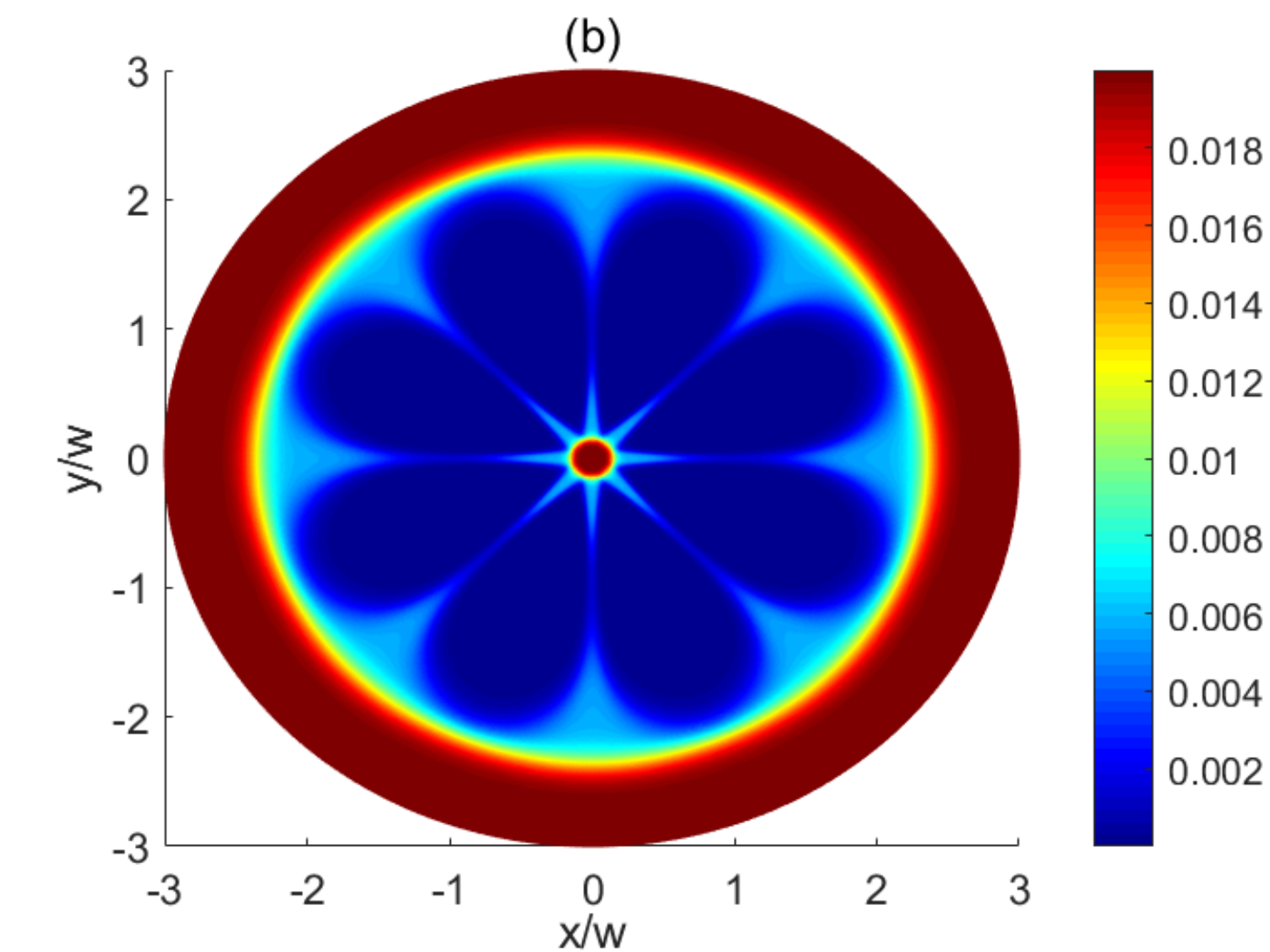}
\includegraphics[width=0.3\columnwidth]{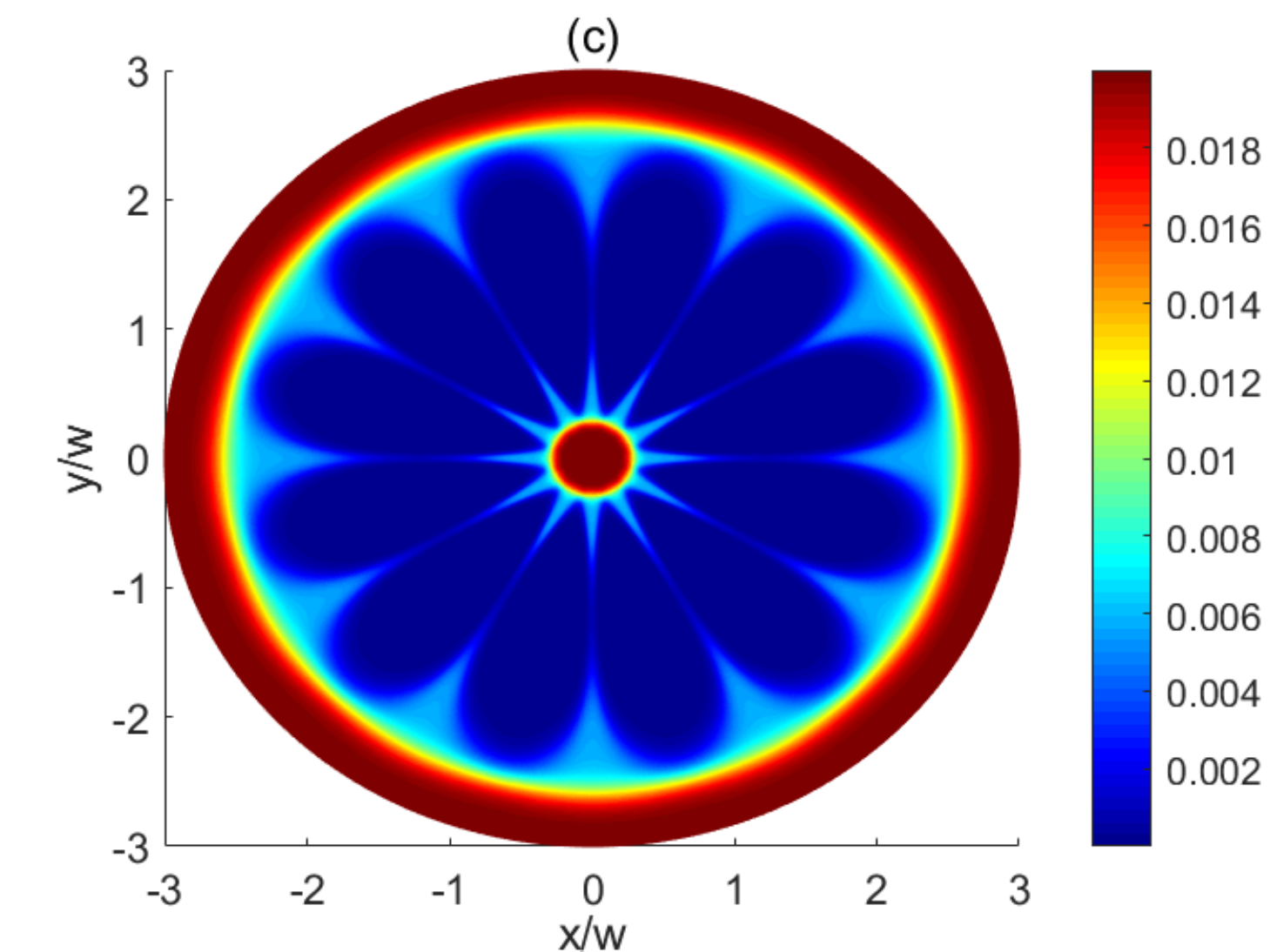}

\includegraphics[width=0.3\columnwidth]{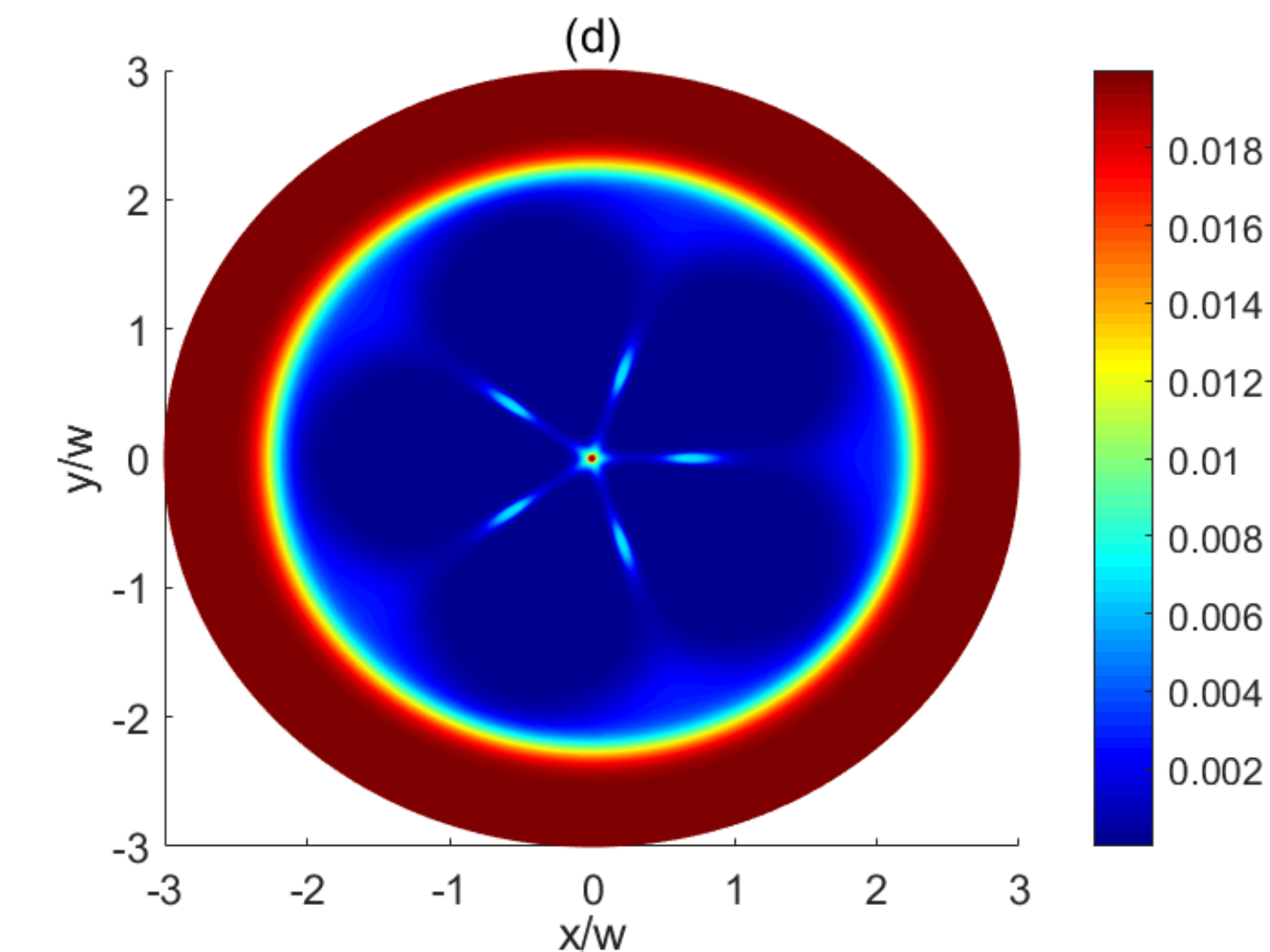} \includegraphics[width=0.3\columnwidth]{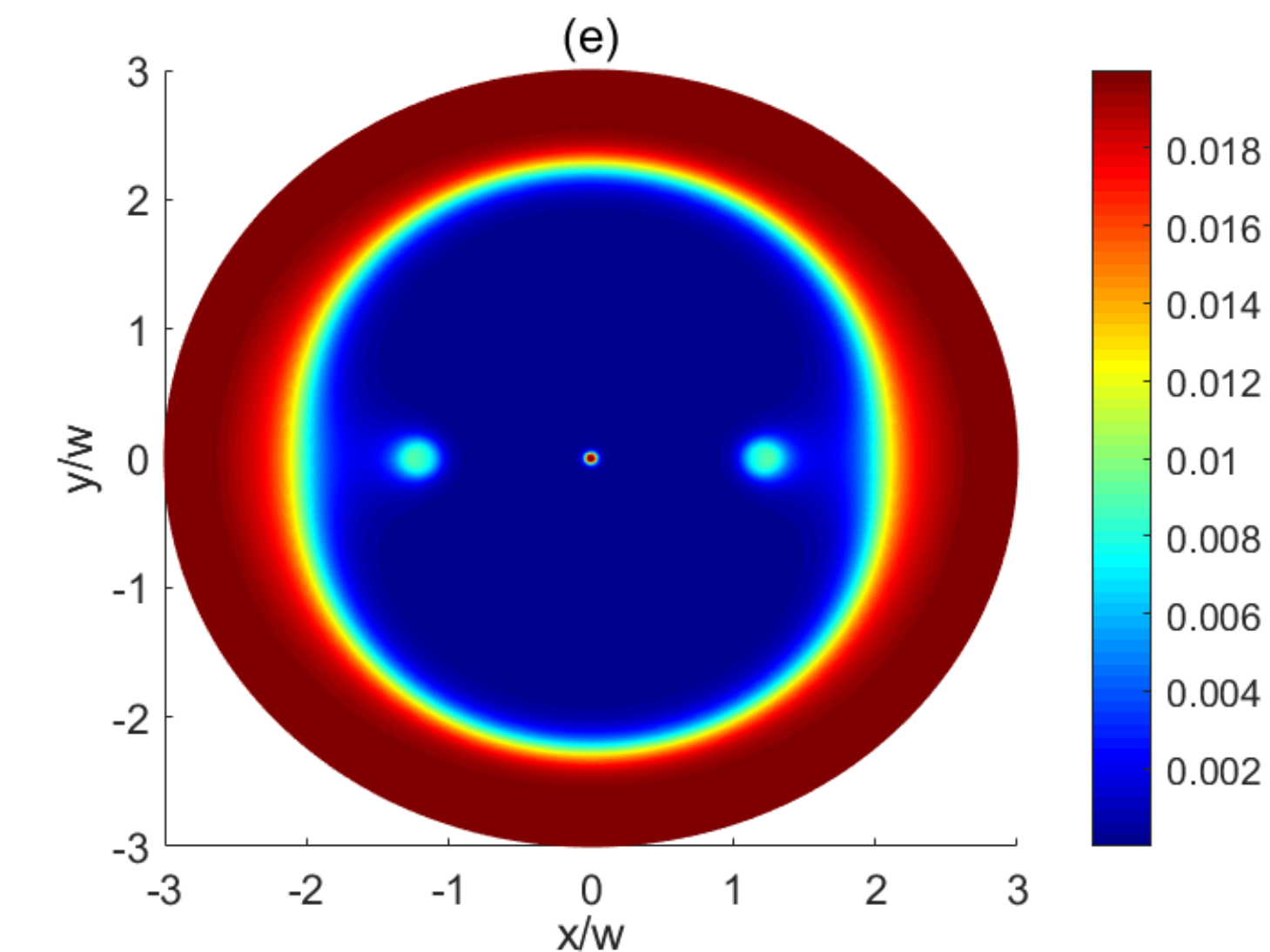}
\caption{Spatially dependent absorption profile of the probe beam in arbitrary
units when all control fields are vortex beams. The vorticities are
$l=1$ (a), $l=2$ (b), $l=3$ (c), $l_{1}=-2$, $l_{2}=1$, $l_{3}=1$,
$l_{4}=-1$ (d) and $l_{1}=1$, $l_{2}=2$, $l_{3}=4$, $l_{4}=3$
(e). The selected parameters are $\epsilon_{1}=0.6\Gamma,$ $\epsilon_{2}=0.7\Gamma,$
$\epsilon_{3}=0.3\Gamma,$ $\epsilon_{4}=0.5\Gamma,$ and the other
parameters are the same as Fig.~\ref{fig:fig3}.}
\label{fig:fig8}
\end{figure}

When $l_{2}=l_{3}=-l_{1}=-l_{4}=l$, Eq.~\ref{eq:QI6} simplifies 

\begin{equation}
Q=2|\Omega_{1}||\Omega_{2}||\Omega_{3}||\Omega_{4}|\cos\left(4\Phi\right),\label{eq:QI6-1}
\end{equation}
and absorption profile shows a $4$-fold symmetry, as illustrated
in Figs.~\ref{fig:fig8}(a)\textendash (c). For different optical
vortices, Eq.~\ref{eq:QI6} results in a $l_{2}+l_{3}-l_{1}-l_{4}$-symmetry
of absorption profile. For example, for $l_{1}=-2$, $l_{2}=1$,$l_{3}=1$,
$l_{4}=-1$ we observe a $5$-fold symmetry (Fig.~\ref{fig:fig8}(d)),
while for $l_{1}=1$, $l_{2}=2$,$l_{3}=4$, $l_{4}=3$ the absorption
profile exhibits a $2-$fold symmetry (Fig.~\ref{fig:fig8}(e)). 

\section{Concluding Remarks \label{sec:conc}}

In conclusion, we have theoretically investigated the spatially structured
optical transparency in a five-level CTL atom-light coupling scheme
illuminated by a weak nonvortex probe beam as well as control laser
fields of larger intensity which can carry OAMs. As a result of the
closed-loop structure of CTL scheme, the linear susceptibility of
the weak probe beam depends on the azimuthal angle and OAM of the
control beams. Therefore, it is possible to obtain information about
positions of low light transmission or optical transparency through
measuring the resulting absorption spectra. Different situations where
one or a combination of control fields are vortex beams are considered,
and different symmetry situations for the absorption profile are obtained.
It is shown that the the quantum interference parameter $Q$ featured
in Eq.~(\ref{eq:QIT}) governs the symmetry of absorption profiles.
In particular, the absorption image of probe field shows $l$-, $2l$-
$3l$-, and even $4l$-fold symmetries due to $l$-, $2l$- $3l$-
or $4l$-fold cosinusoidal behavior of the parameter $Q$, defined
by Eq.~(\ref{eq:QIT}). The spatially varying optical transparency
may find potential applications in storage of high-dimensional optical
information in phase dependent quantum memories.

A possible realistic experimental realization of the proposed combined
tripod and $\Lambda$ setup can be implemented e.g. for the Cs atoms.
The lower levels $|a\rangle$, $|c\rangle$ and $|d\rangle$ can be
assigned to $|6S_{1/2},F=4\rangle$, $|6S_{1/2},F=3,M_{F}=+1\rangle$
and $|6S_{1/2},F=3,M_{F}=+3\rangle$, respectively. Two excited states
$|b\rangle$and $|e\rangle$ can be attributed to $|6P_{3/2},F=4\rangle$
and $|6P_{3/2},F=2,M_{F}=+2\rangle$, respectively. 
\begin{acknowledgments}
This research was funded by the European Social Fund under grant No.\ 09.3.3-LMT-K-712-01-0051. 
\end{acknowledgments}

\end{document}